\documentclass[useAMS,usenatbib,referee]{mn2e}
\usepackage{graphicx}
\usepackage{color}

\usepackage{longtable}
\usepackage{lscape}
\usepackage{subfigure}

\title[Modeling the ISO-LWS high resolution far-IR H$_{2}$O
lines from Orion-KL]
{Chemical and radiative transfer modeling of the
ISO-LWS Fabry-Perot spectra of Orion-KL water lines}
\author[M. R. Lerate et al.]
{M. R. Lerate,$^{1,2}$, J. A. Yates$^1$, M. J. Barlow$^1$, S. Viti$^1$, B. M.
Swinyard$^3$\\
  $^1$University College London, Gower Street, London WC1E 6BT,
  U.K.\\
$^2$ European Space Astronomy Center ESAC, Villanueva de la Cañada, 28691, Madrid, Spain.\\
  $^3$Rutherford Appleton Laboratory, Chilton, Didcot OX11 0QX,
  U.K.}

\date{2009 Xxxxx XX}

\pagerange{\pageref{firstpage}--\pageref{lastpage}} \pubyear{2009}

\def\LaTeX{L\kern-.36em\raise.3ex\hbox{a}\kern-.15em
    T\kern-.1667em\lower.7ex\hbox{E}\kern-.125emX}

\begin{document}

\label{firstpage}

\maketitle

   \begin{abstract}

{We present chemical and radiative transfer models for the many far-IR
ortho- and para-H$_{2}$O lines that were observed from the Orion-KL region
in high resolution Fabry-P\'erot (FP) mode by the Long Wavelength
Spectrometer (LWS) on board the {\em Infrared Space Observatory (ISO)}.
The chemistry of the region was first studied by simulating the conditions
in the different known components of Orion-KL: chemical models for a hot
core, a plateau and a ridge were coupled with an accelerated
$\Lambda$-iteration (ALI) radiative transfer model to predict H$_2$O line
fluxes and profiles. Our models include the first 45 energy levels
of ortho- and para-H$_{2}$O. We find that lines arising from
energy levels below 560~K were best reproduced by a gas of density
3$\times$10$^{5}$~cm$^{-3}$ at a temperature of 70-90~K, expanding at a
velocity of 30~km~s$^{-1}$ and with a H$_{2}$O/H$_{2}$ abundance ratio of
the order of 2 - 3 $\times$ 10$^{-5}$, similar to the abundance derived by
Cernicharo et al. (2006). However, the model that best reproduces the
fluxes and profiles of H$_{2}$O lines arising from energy levels above
560~K has a significantly higher H$_{2}$O/H$_{2}$ abundance, 1 - 5
$\times$
10$^{-4}$, originating from gas of similar density, in the Plateau region,
that has been heated to 300 K, relaxing to 90-100 K. We
conclude that the observed water lines do not originate from
high temperature shocks.}

 \end{abstract}


\section{Introduction}
High velocity gas was first detected at the centre of the Orion-KL
region as broad wings on `thermal' molecular lines in the
millimeter range and as high velocity maser features in the 22 GHz
line of H$_{2}$O (Genzel et al. 1981). These high velocity motions
may be caused by mass outflows from newly formed stars. Many
theoretical studies of the Orion region (e.g. Draine \& Roberge
1982; Chernoff et all. 1982; Neufeld \& Melnick 1987) have
concluded that the rich emission spectrum of thermally excited
water vapor should play a
significant role in cooling the gas. \\

The widespread distribution of water vapour around IRc2 has been
probed with maps at 183 GHz (Cernicharo et al. 1990, 1994,
Cernicharo \& Crovisier 2005), the first time that its abundance
was estimated in the different large-scale components of Orion
IRc2. Harwit et al. (1998) analysed 8 water lines observed with the ISO
LWS FP, concluding that these lines arise from a molecular cloud
subjected to a magnetohydrodynamic C-type shock. From their
modelling, they derived an H$_{2}$O to H$_2$ abundance ratio of
5$\times$ 10$^{-4}$. However, the interpretation of the lines observed
in the $\approx$ 80$^{\prime\prime}$ LWS beam and the determination of
the water abundance in the different Orion components remains a
long standing problem, due to two main issues:
the complexity of the dynamical and chemical processes
that are taking place within the region encompassed by the LWS beam,
with outflows and several different gas components, and the need for
H$_{2}$O collisional rates appropriate for the temperatures
prevailing in shocks. \\

A total of 70 water lines were detected in the ISO-LWS far-IR
spectral survey of Orion-KL (Lerate
et al. 2006), with typically 70 km s$^{-1}$ FWHM.
The line profiles range from
predominantly P-Cygni at shorter wavelengths to predominantly pure
emission at longer wavelengths. At the shorter wavelengths, the
heliocentric velocities of absorption components appear to be
centred at $\approx$ -- 15 km s$^{-1}$,
consistent with the results found in the ISO
Short Wavelength Spectrometer (SWS) wavelength range, shortwards of
45~$\mu$m (van Dishoeck et al. 1998; Wright et al. 2000). However,
the radial velocities of the pure emission lines of H$_{2}$O peak at
around +30~km~s$^{-1}$, whereas the velocity of the Orion-KL quiescent
gas is +9~km~s$^{-1}$. \\

In a previous analysis of the H$_2$O lines observed by the ISO
LWS-FP, Cernicharo et al. (2006) modeled lines from the first 30
rotational levels of ortho- an para-H$_2$O, concluding that the
lines mainly arise from a gas flow expanding at 25-30 km s$^{-1}$,
and inferred a gas temperature of approximately 80-100 K and a
H$_{2}$O/H$_{2}$ abundance ratio of 2--3 $\times$ 10$^{-5}$. This
derived abundance was an average over the LWS beam and they
suggested that water could be locally more abundant if the
emitting region included
warmer components which interact with the ambient gas. \\

In the current work, a technique to
distinguish between the different
components has been applied to the final calibrated high-resolution
spectra of the ortho- and para-H$_{2}$O lines from the Orion-KL region.
The methodology is similar to that used to model the
ISO-LWS high resolution spectra of the Orion-KL CO lines (Lerate et al.
2008).
Chemical models of the physically distinct components are coupled
with a non-local radiative transfer model. The description of the
models is structured to follow the different components
of the region: the hot core, plateau and ridge. Many references
can be found for a description of the KL region components - a
complete overview is given in Stahler and Palla (2004). Both
models (chemical and radiative transfer) are highly configurable
and have been applied to a variety of emitting regions,
including molecular gas in outflows (Benedettini et al. 2006).

\section{Observations and data reduction}
The ISO-LWS FP observations were carried out between September 1997 and
April
1998. The dataset consists of 26 individual observations making up
a total of 27.9 hours of {\em ISO} LWS observing time in L03
Fabry-Perot full spectral scan mode. The dataset also
includes 16 observations accumulated over 13.1 hours in L04
Fabry-Perot line scan mode and 1
observation in the lower resolution L01 grating scan mode. The
instrumental field of view for all L03 observations was centred
either on a position offset by 10.5$^{\prime\prime}$ from the BN
object (which is at 05$^{\rm h}$ 35$^{\rm m}$ 14.12$^{\rm s}$ --
05$^{\circ}$ 22$^{\prime}$ 22.9$^{\prime\prime}$ J2000), or on a
position offset by 5.4$^{\prime\prime}$ from IRc2 (which is at
05$^{\rm h}$ 35$^{\rm m}$ 14.45$^{\rm s}$ -- 05$^{\circ}$
22$^{\prime}$ 30.0$^{\prime\prime}$ J2000), while most of the L04
observations were centred on IRc2. The LWS beam had a diameter
$\approx$ 80$^{\prime\prime}$ (Gry et al. 2003). Processing of the
LWS FP data was carried out using the Offline Processing (OLP)
pipeline and the LWS Interactive Analysis (LIA) package version
10. The basic calibration is fully described in the LWS handbook
(Gry et al. 2003). Further processing, including dark current
optimisation, de-glitching and removal of the LWS grating profile
was then carried out interactively using the LWS Interactive Analysis
package version 10
(LIA10; Lim et al. 2002) and the {\em ISO} Spectral Analysis
Package (ISAP; Sturm 1998). A detailed description of the observations
and of the data reduction process is given by Lerate et al.
(2006).

\section{The chemical and radiative transfer models}
\subsection{The Chemical Models}
The chemical model used to simulate the Orion KL plateau and ridge
is described by Viti et al. (2004) and is the same as used by
Lerate et al. (2008) to model the Orion-KL lines observed by the
ISO-LWS Fabry-Perot. The chemical network is taken from the UMIST
database (Le Teuff et al. 2000). The model uses a two phase
calculation in which gravitational collapse occurs in phase I,
with gas-phase chemistry and sticking onto dust particles with
subsequent processing (hydrogenation and conversion of a fraction
of CO into methanol) also occurring. In phase II, where
evaporation from grains also took place, we simulated the presence
of a non-dissociative shock by an increase of temperature (from
200 to 2000K, depending on the model) at an age of $\approx$ 1000
yr, which is the dynamical timescale of the main outflow observed
in the KL region (Cernicharo et al. 2006). The duration of the
high temperature shock is about 100 years, representing the
temperature structure of a C-shock. The gas is then allowed to
cool. This temperature profile was adopted from the calculation of
Bergin et al. (1998) who studied the chemistry of H$_2$O and O$_2$
in postshock gas. The treatment of the temperature increase is
considered to be linear with time. The modeling allows a wide
range of parameters to be varied in order to investigate a range
of conditions. As with the CO modelling of Lerate et al. (2008),
the main parameters that were varied in this analysis were: (i)
the initial and final densities (ii) the depletion efficiency,
hereafter called the freez-out parameter (iii) the cloud size (iv)
the maximum gas temperature and (v) the interstellar radiation
field (ISRF). Note that the H$_2$ column density is not a free
input parameter but is calculated
self-consistently as a function of size and gas density.  \\

We based the parameter grid on descriptions of the Orion-KL
components found in detailed analyses of the region (Blake et al.
1987, Genzel and Stutzki 1989, Cernicharo et al. 1994). The main
parameters adopted from these references were the sizes and
densities. However, we used our own analysis of the continuum
emission and molecular rotational diagrams (from Lerate et al.
2006) to indicate gas temperatures. Table~\ref{plat_para} list the
main parameter sets used to compute the grid of chemical models
that were used for the present H$_2$O modelling and for the CO
modelling of Lerate et al. (2008). For completeness we also list
the derived H$_2$ column densities for each model.

\begin{table*}
\centering
\begin{tabular}{ c c c c c c c}
\hline\hline  Model &Density (cm$^{-3}$) & T$_{shock}$ (K)&
T$_{gas}$ (K) &mco\% & size ($\prime\prime$) & N($H_2$)(cm$^{-2}$)  \\
\hline
PL1&3$\times$10$^{5}$ & 200 & 80 & 40 & 30 & 4$\times$10$^{22}$\\
PL2&3$\times$10$^{5}$ & 300 & 90 & 60 & 30 & 4$\times$10$^{22}$\\
PL3&3$\times$10$^{5}$ & 500 & 90 & 60 & 30& 4$\times$10$^{22}$\\
PL4&3$\times$10$^{5}$ & 500 & 90 & 80  & 30& 4$\times$10$^{22}$\\
PL5&3$\times$10$^{5}$ & 1000 & 100 & 60  & 30& 4$\times$10$^{22}$ \\
PL6&3$\times$10$^{5}$ & 1000 & 100 & 80  & 30& 4$\times$10$^{22}$\\
PL7&3$\times$10$^{5}$ & 2000 & 100 & 60  & 30& 4$\times$10$^{22}$\\
PL8&3$\times$10$^{5}$ & 2000 & 100 & 80 & 30& 4$\times$10$^{22}$\\
PL9&1$\times$10$^{6}$ & 300 & 90 & 60 & 30& 2$\times$10$^{23}$\\
PL10&1$\times$10$^{6}$ & 300 & 90 & 80& 30& 2$\times$10$^{23}$\\
PL11&1$\times$10$^{6}$ & 500 & 90 & 60& 30& 2$\times$10$^{23}$\\
PL12&1$\times$10$^{6}$ & 500 & 90 & 80 & 30& 2$\times$10$^{23}$\\
PL13&1$\times$10$^{6}$ & 1000 & 100 &60& 30& 2$\times$10$^{23}$\\
PL14&1$\times$10$^{6}$ & 1000 & 100 &80& 30& 2$\times$10$^{23}$\\
PL15&1$\times$10$^{6}$ & 2000 & 100 &60& 30& 2$\times$10$^{23}$\\
PL16&1$\times$10$^{6}$ & 2000 & 100 &80& 30& 2$\times$10$^{23}$\\
\hline
RG1&1$\times$10$^{4}$ & no shock & 70 & 40 & 15 & 9$\times$10$^{20}$\\
RG2&1$\times$10$^{4}$ & no shock & 70 & 20 & 15 & 9$\times$10$^{20}$\\
RG3&5$\times$10$^{4}$ & no shock & 80 & 40 & 15 & 5$\times$10$^{21}$\\
RG4&5$\times$10$^{4}$ & no shock & 80 & 20 & 15  & 5$\times$10$^{21}$\\
 \hline
 \end{tabular}
 \caption{Plateau (PL) and Ridge (RG) models and their parameters: density, maximum temperature reached by the gas in the shock simulation, gas
temperature after cooling, the percentage of mantle CO (mco) given by the freeze-out parameter
at the end of Stage I of the chemical model, and the approximate H$_2$
column density (assuming $\sim$ 1.6$\times$10$^{21}$ cm$^{-2}$ per
magnitude of visual extinction).}
 \label{plat_para}
 \end{table*}

The chemical model used to simulate the extended gas in
Orion KL is based on the same model used for the shocked gas, with
the exception that no high temperature shock is included as an input.
The model simulates the gas chemistry evolution of gas expanding
with velocities of $\approx$ 25-30 km s$^{-1}$, which is heated up
to $\approx$ 100 K on a timescale of 1000 yr, relaxing to 80 K
(see Figure~\ref{extended}). A grid of conditions was also
investigated for six extended-gas models (Table ~\ref{exten_para}).
Their initial parameters
were based on the published literature on the Orion-KL extended gas
(Tielens and Hollenbach (1985), Draine and Roberge (1982), Menten et al.
(1990)).\\

\begin{table}
\centering
\begin{tabular}{ c c c c c c }
\hline\hline  Model &Density (cm$^{-3}$) &
T$_{gas}$ (K) &mco\% & expansion gas velocity (km s$^{-1}$) \\
\hline

E1&3$\times$10$^{5}$  & 80 & 80 & 25\\
E2&3$\times$10$^{5}$  & 90 & 70 & 30 \\
E3&5$\times$10$^{4}$  & 100 & 60 & 35 \\
E4&3$\times$10$^{5}$  & 80 & 50 & 25\\
E5&3$\times$10$^{5}$  & 90 & 40 & 30 \\
E6&5$\times$10$^{4}$  & 100 & 30 & 35\\
 \hline
 \end{tabular}
 \caption{List of main parameters modified to investigate the extended-gas chemical model of Orion KL.}
 \label{exten_para}
 \end{table}

\subsection{The ortho-H$_{2}$O and para-H$_{2}$O radiative transfer
models}

As described by Lerate et al. (2008), the chemical model produces
abundances that are used as inputs to the radiative transfer model
SMMOL (Rawlings \& Yates, 2001), along with values for the
dimensions, density, dust temperature and  radiation field
intensity. The SMMOL code has accelerated $\Lambda$-iteration
(ALI) that solves the radiative transfer problem in multi-level
non-local conditions. It starts by calculating the level
population assuming LTE and takes an adopted radiation field as
the input continuum and then recalculates the total radiation
field and level populations, repeating the process until
convergence is achieved. The resulting emergent intensity
distribution is then transformed to the LWS units of flux (W
cm$^{-2}$ $\mu$m$^{-1}$) taking into account the different beam
sizes (slightly different for each detector) and is convolved with
an instrumental line profile corresponding to a 33 km s$^{-1}$
FWHM Lorentzian (Polehampton et al. 2007), in order to directly
compare with the observations. The input parameters include
molecular data such as the molecular mass, energy levels,
radiative and collisional rates and also the dust size
distribution and opacity. The parameters also include the gas and
dust temperatures of the object being modelled. We estimated the
dust temperature from the LWS grating observation of Orion-KL
(Lerate et al. 2006), fitting a black body function of 70 K. In
order to reproduce the observed continuum flux level, we adopted a
radiation field equivalent to 10$^{4}$ Habings (1 Habing unit
corresponds to the integrated flux in the wavelength range from
91.2 to 111.0 nm of 1.6 $\times$ 10$^{-3}$ erg s$^{-1}$ cm$^{-2}$;
Habing 1968). The adopted ortho-para ratio was 3, which was found
by Barber et al. (2006) to be appropriate when T$>$50~K. The
microturbulent velocity was set to 5 km s$^{-1}$ and expansion
velocities from 15--40 km s$^{-1}$ were considered for the shocked
gas in the Plateau models. We estimated an error of less than 30\%
for the fit to the continuum, being the maximum percentage
deviation below the observed continuum for wavelengths up to
120~$\mu$m and the maximum percentage deviation above the observed
continuum for longer wavelengths.

The o-H$_{2}$O molecular data were taken from public molecular
databases (M\"{u}ller et al., 2001; Sch\"{o}ier et al., 2005).
Table \ref{oh2o_inputs} lists the input parameters for the
radiative transfer model that were modified with respect to the CO
modelling. The number of energy levels was set to 45 and the
number of radiative transitions to 158. The number of collisional
transitions was 990 and 10 collisional temperatures were
investigated, from 20 K to 2000 K. The H$_2$O-H$_2$ collisional
excitation rates were based on the (scaled) H$_2$O-He calculations
of Green et al. (1993). Note that since the calcuations for this
paper were performed new collisional rates for H$_2$O-H$_2$ have
become available (Faure et al. 2007; Dubernet et al. 2009).
Table 1 of Faure et al. shows that for temperatures below 300K the
critical densities with helium can be factors of 2-5 greater than
for rates with ortho and para hydrogen. However as our densities
are at least 3 orders of magnitude below the reported critical
densities, the radiative pumping terms are as important, if not
more so, than the collisional pumping terms. We would therefore
expect our results to be qualitatively valid within the
observational errors.  However one would expect the use of He as a
collisional partner to be incorrect when trying to model Herschel
HIFI data, because the much better angular and spectral resolution
of the observations may reveal pockets of much higher density; in
such cases the use of the new collisional rates for water will be
necessary.

\begin{table}
\scriptsize \centering
\begin{tabular}{ c c }
\hline\hline
\textbf{INPUT PARAMETERS} &\textbf{VALUE} \\
\hline \hline
MOLECULAR WEIGHT & 18.0\\
\hline
NUMBER OF RADIATIVE TRANSITIONS & 158 \\
\hline
NUMBER OF COLLISION PARTNERS & 1\\
\hline
NUMBER OF COLLISION TEMPERATURES & 10 \\
\hline
NUMBER OF COLLISIONAL TRANSITIONS & 990\\
\hline

ISUM: THE STATISTICAL EQUILIBRIUM EQUATION FOR LEVEL ISUM IS & 45\\
REPLACED BY THE EQUATION OF CONSERVATION OF NUMBERS& \\
\hline
NK: NUMBER OF ENERGY LEVELS INCLUDING CONTINUUM LEVELS & 45\\
\hline
 \end{tabular}
 \caption{List of main o-H$_{2}$O and  p-H$_{2}$O parameters adopted as
inputs in the radiative transfer models}
 \label{oh2o_inputs}
 \end{table}


\section{Results and Discussion}

\begin{figure}
    \centering
\includegraphics[width=10cm,height=10cm]{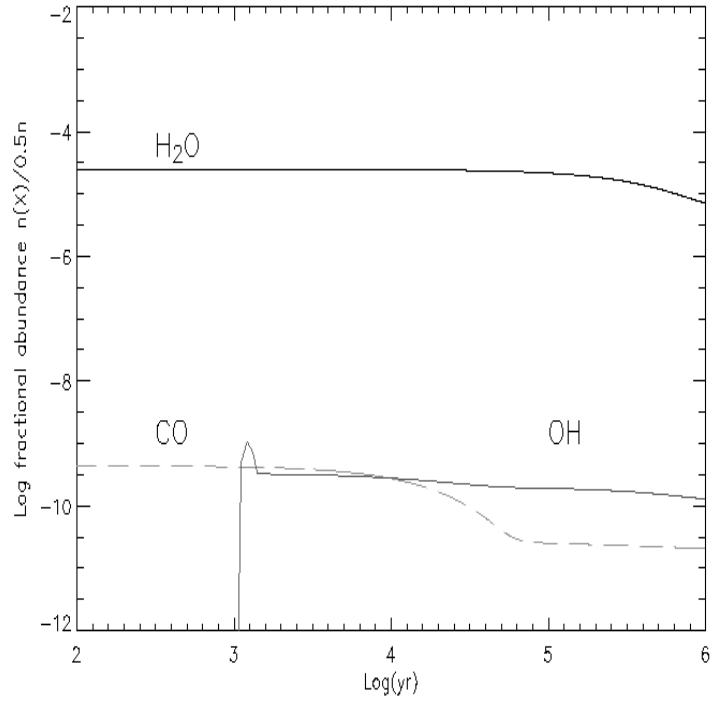}
    \caption{Time evolution H$_{2}$O, OH and CO abundances corresponding to the
     extended gas chemical model E2 in Table \ref{exten_para}.}
        \label{extended}
   \end{figure}
\begin{figure}
    \centering
\includegraphics[width=10cm,height=10cm]{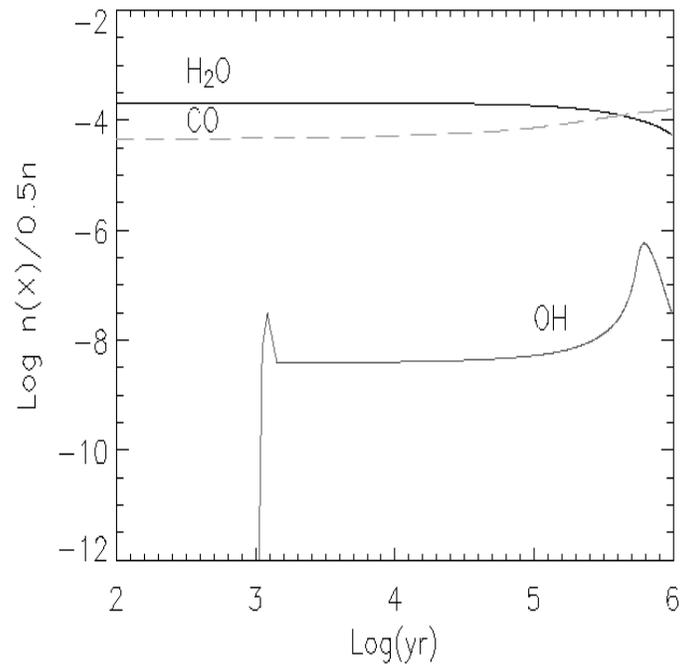}
    \caption{Time evolution of H$_{2}$O, OH and CO abundances for the
     PL2 chemical model (see Table~\ref{plat_para} for a description of the main model parameters).}
        \label{PL2}
   \end{figure}

The H$_2$O line profiles and fluxes were reproduced by coupling our
chemical models (see
Figure~\ref{extended} and Figure~\ref{PL2} for the final adopted
chemical models) with SMMOL radiative transfer models. A large number of
models were investigated, with simulations of different shock
temperatures, see Table \ref{plat_para} and Table
\ref{exten_para}. In the models where the H$_2$O lines were
reasonably well reproduced, an
extended grid of conditions was then investigated to refine the line
fits. An example of this is shown for the Extended gas model, whose
investigated scenarios are summarised in Table \ref{grid_ext}. The
main parameters that were modified to refine the models were the
Phase~II gas temperature and the freeze-out parameter, which determines
the amount of water frozen onto the grains at the end of phase I of the
chemical model.\\

\begin{table}
\centering
\begin{tabular}{ c c c c c c }
\hline\hline  Model &Density (cm$^{-3}$) &
T$_{gas}$ (K) & gas velocity (km s$^{-1}$) & mH$_{2}$O (\%) & H$_{2}$O/H$_{2}$ abundance \\
\hline

W1&3$\times$10$^{5}$  & 80   & 25 & 70 &2.1 $\times$ 10$^{-5}$ \\
W2&3$\times$10$^{5}$  & 90   & 30 & 60& 2.8 $\times$ 10$^{-5}$ \\
W3&5$\times$10$^{4}$  & 100   & 35 & 50& 4.8 $\times$ 10$^{-5}$ \\
W4&3$\times$10$^{5}$  & 80  & 25& 40&7.5 $\times$ 10$^{-5}$ \\
W5&3$\times$10$^{5}$  & 90   & 30 & 30& 9.8 $\times$ 10$^{-5}$ \\
W6&5$\times$10$^{4}$  & 100  & 35 & 20 & 4.8 $\times$ 10$^{-6}$ \\
 \hline
 \end{tabular}
 \caption{List of main parameters and resulting H$_{2}$O/H$_{2}$ abundances from the chemical models adopted to investigate the extended-gas component in Orion-KL.}
 \label{grid_ext}
 \end{table}

\begin{figure}
    \centering
\includegraphics[width=4cm,height=4cm]{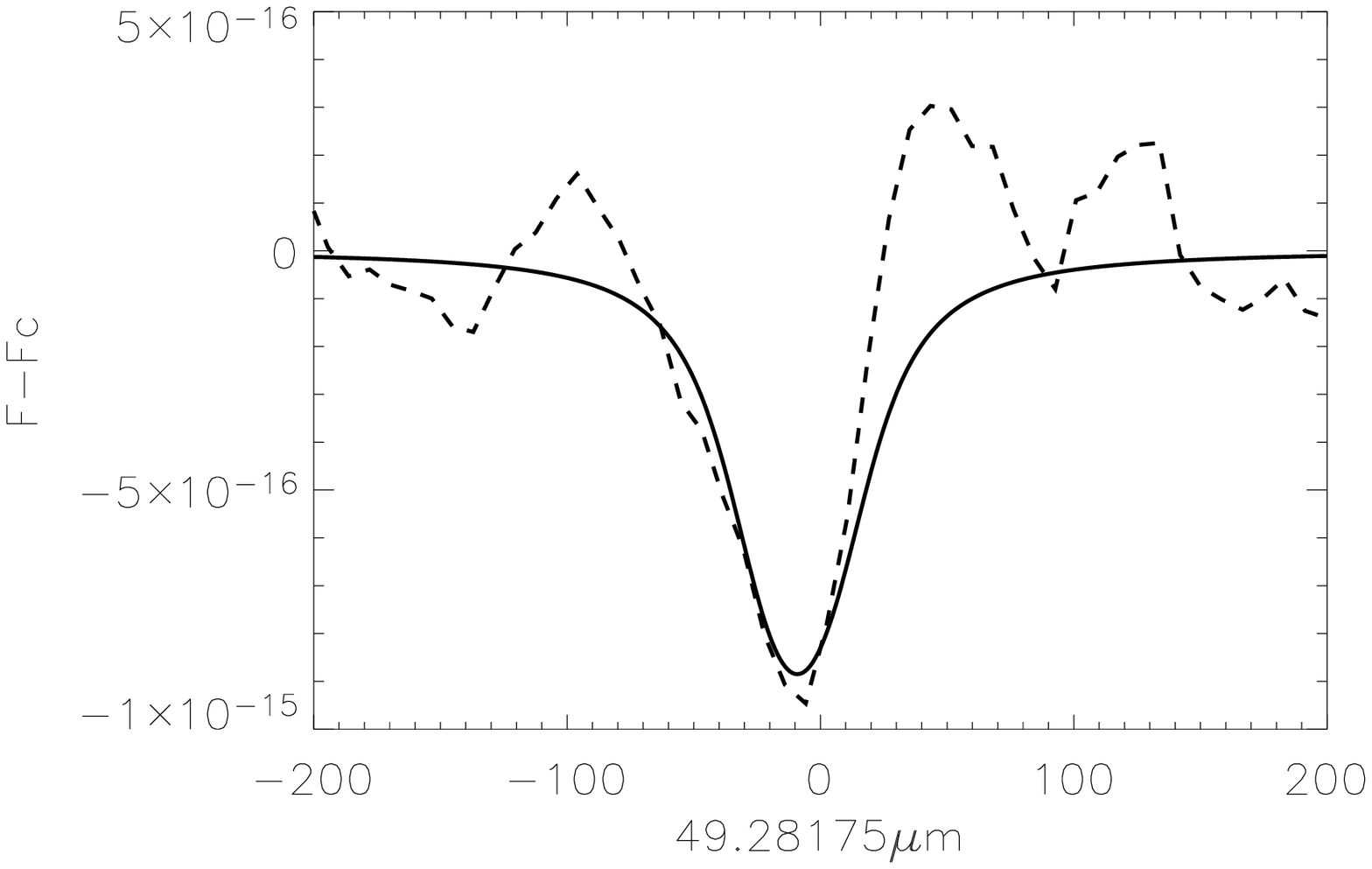}
\includegraphics[width=4cm,height=4cm]{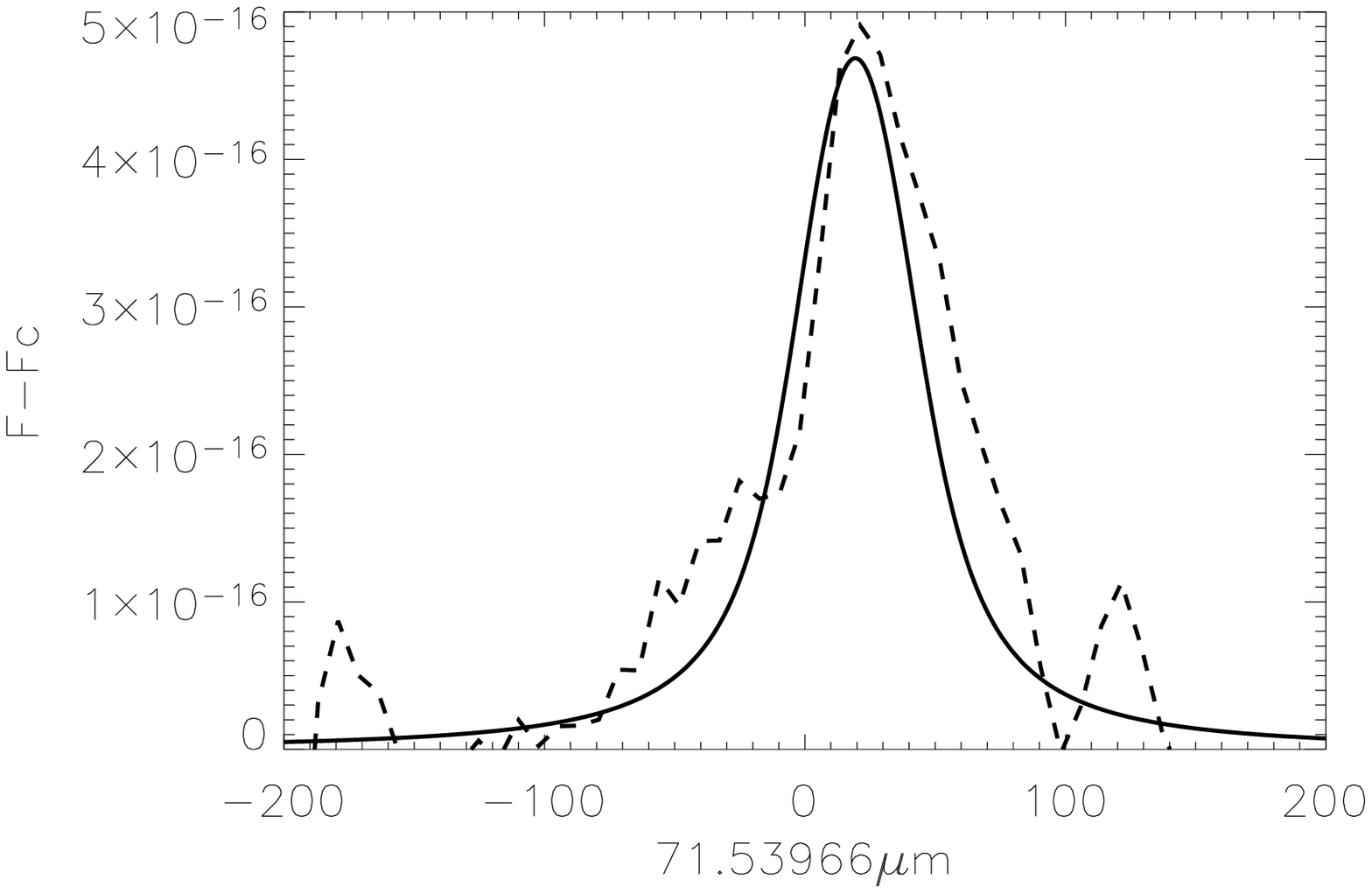}
\includegraphics[width=4cm,height=4cm]{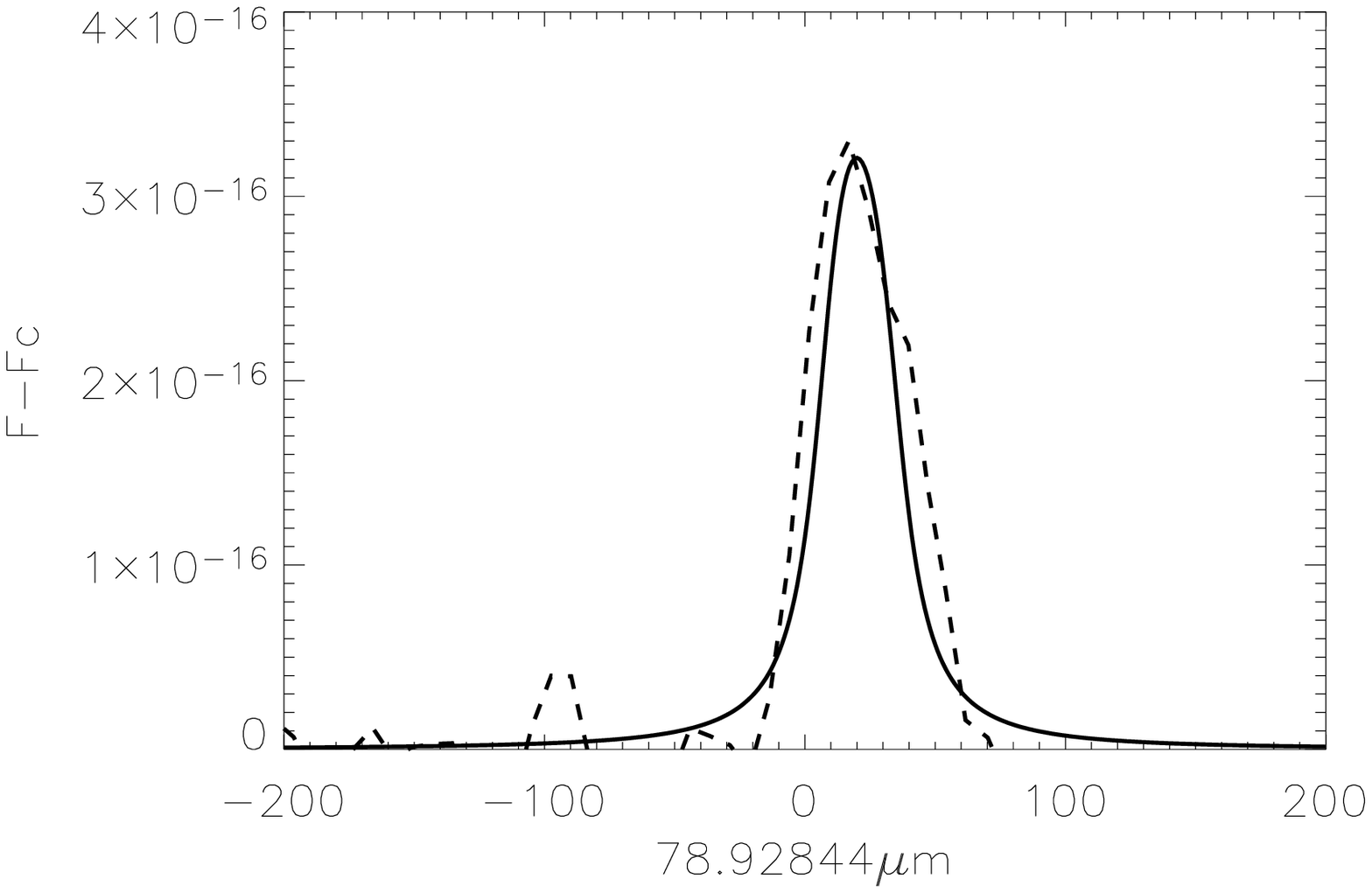}
\includegraphics[width=4cm,height=4cm]{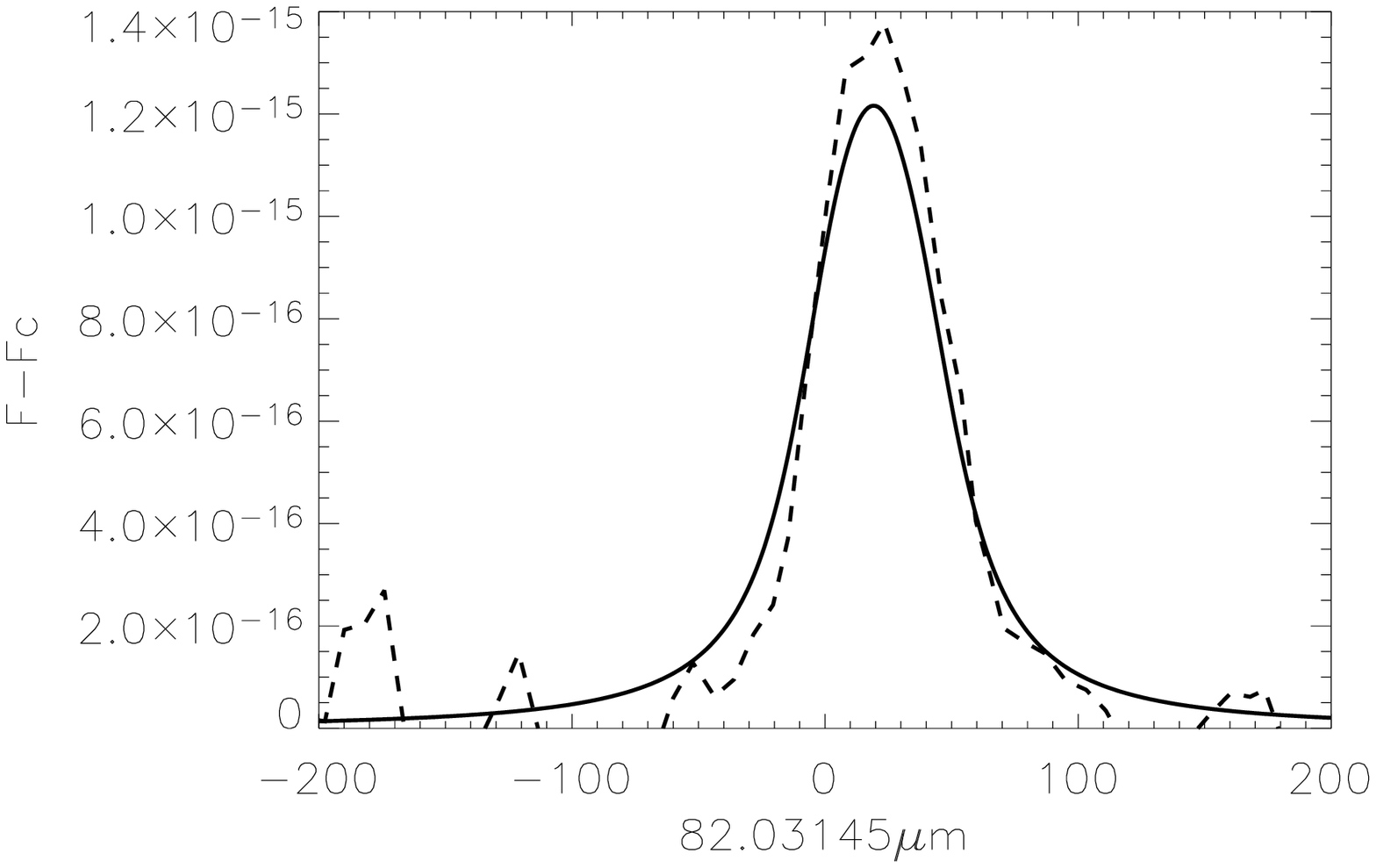}
\includegraphics[width=4cm,height=4cm]{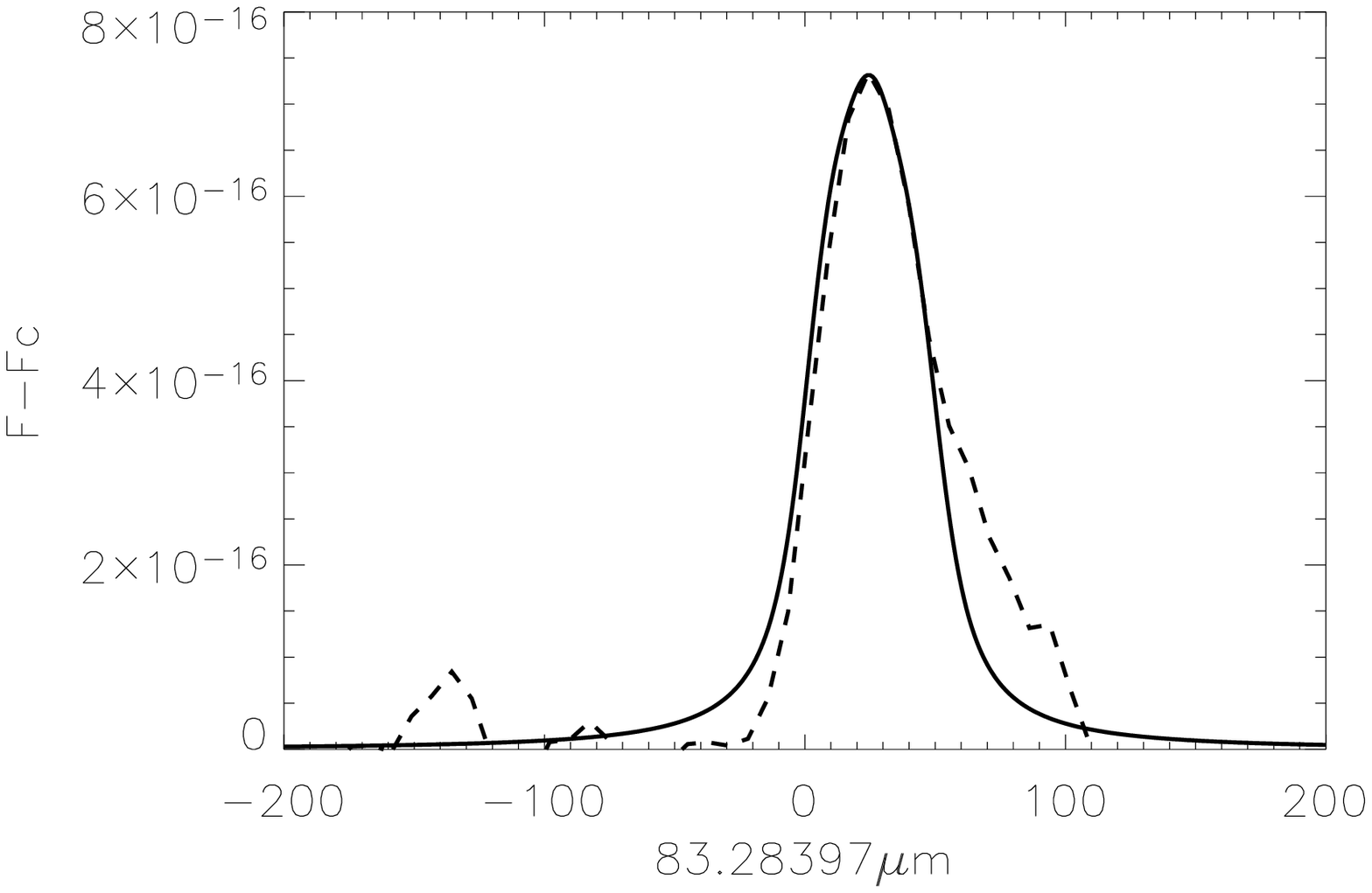}
\includegraphics[width=4cm,height=4cm]{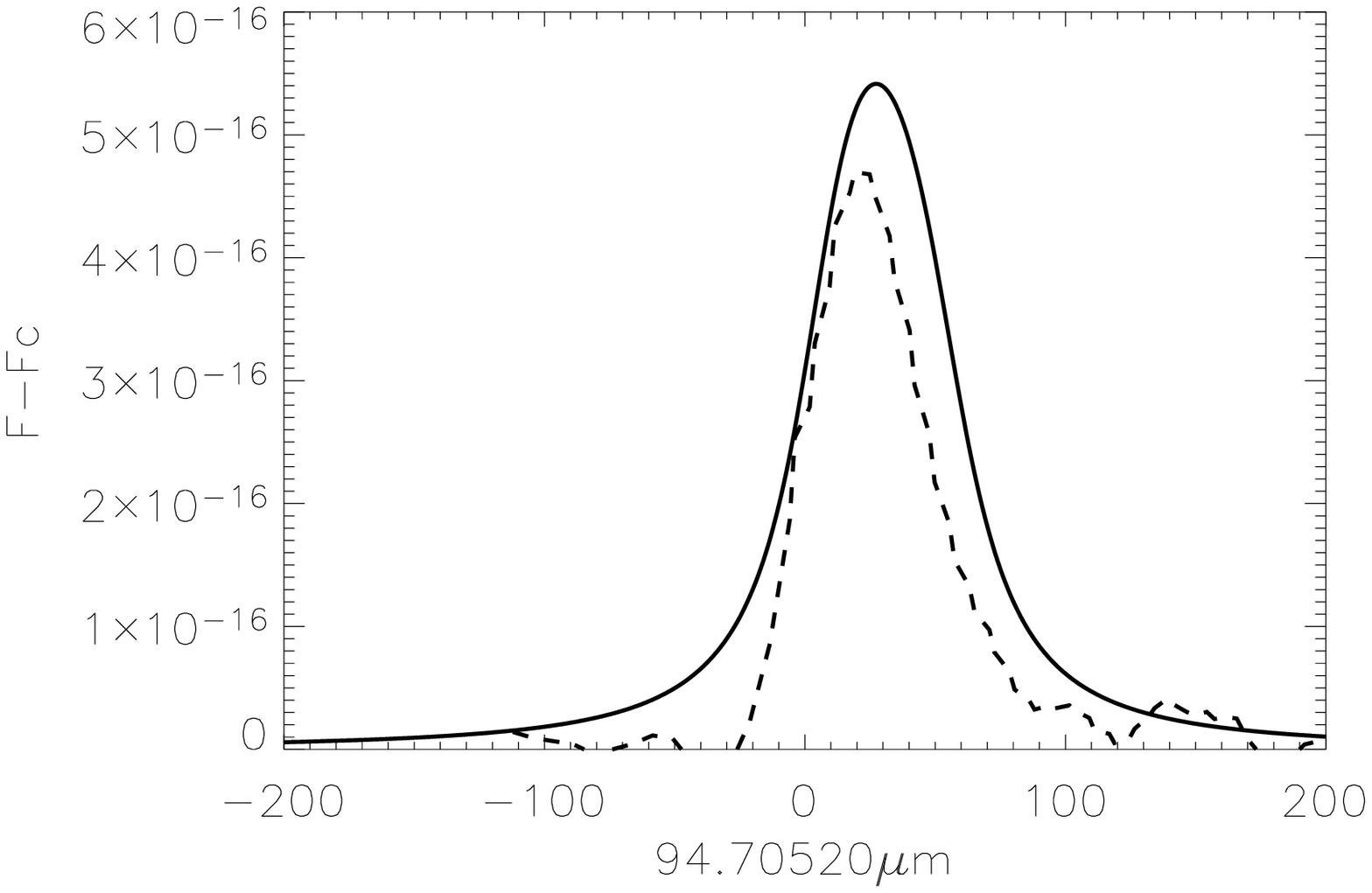}
\includegraphics[width=4cm,height=4cm]{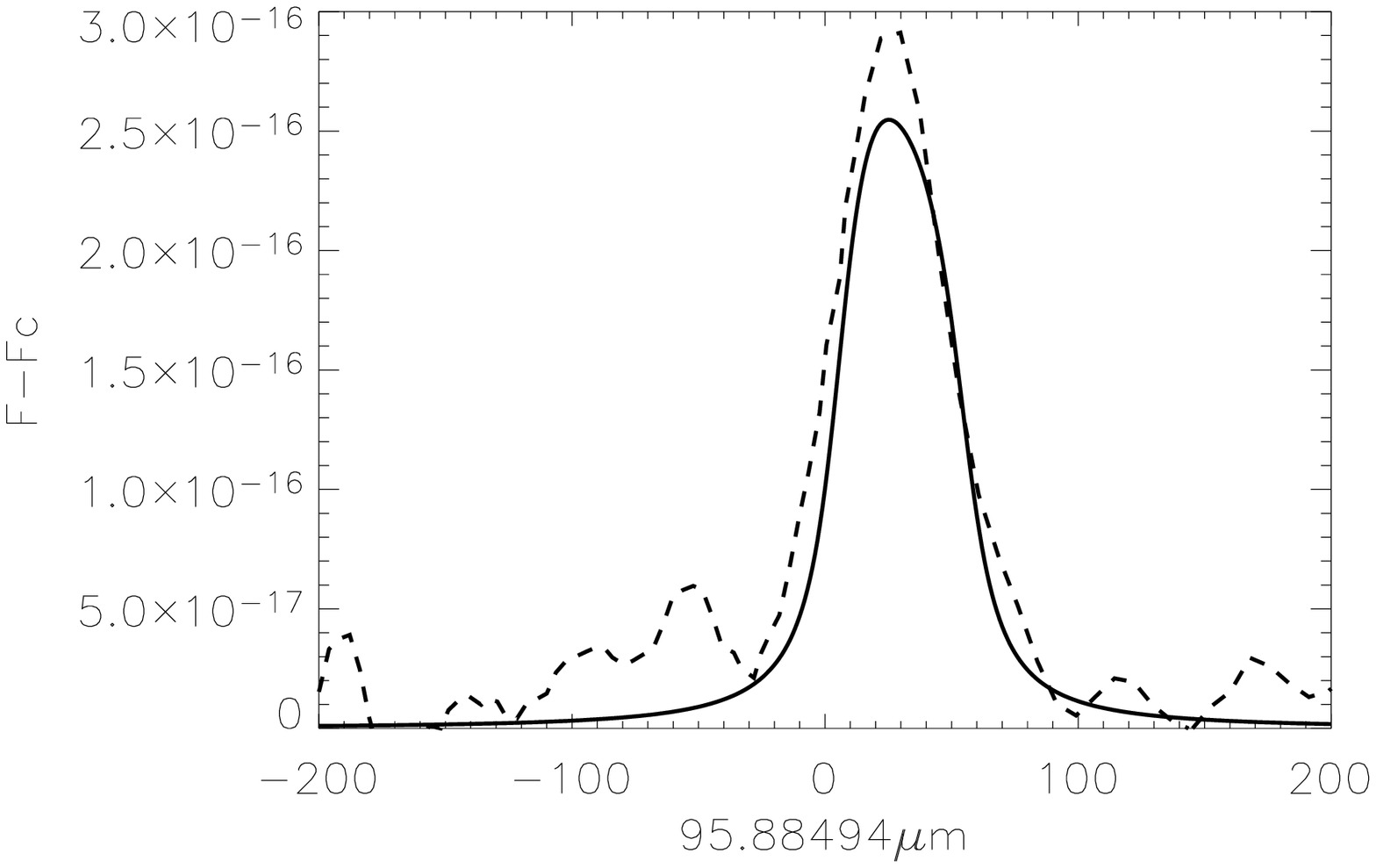}
\includegraphics[width=4cm,height=4cm]{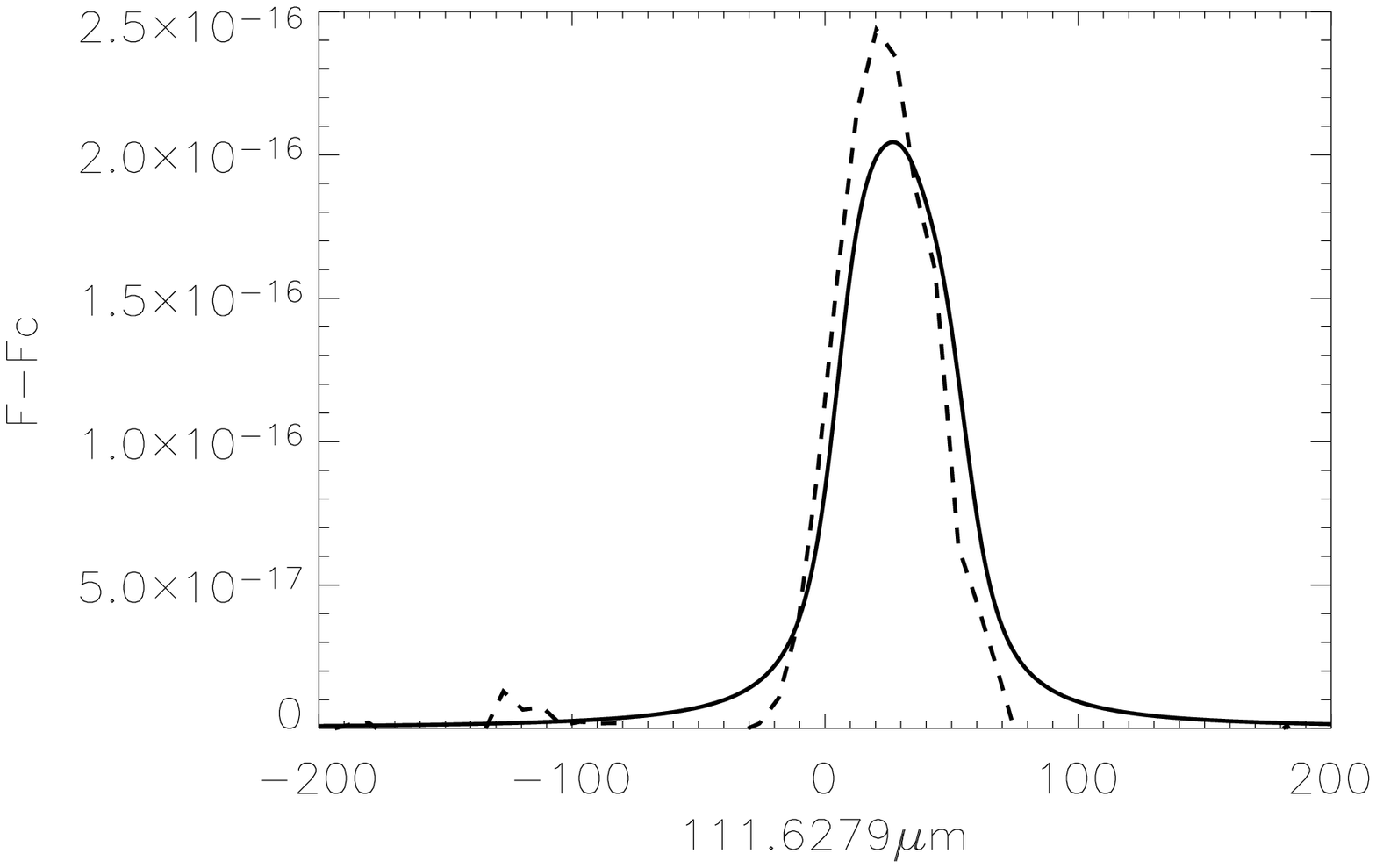}
\includegraphics[width=4cm,height=4cm]{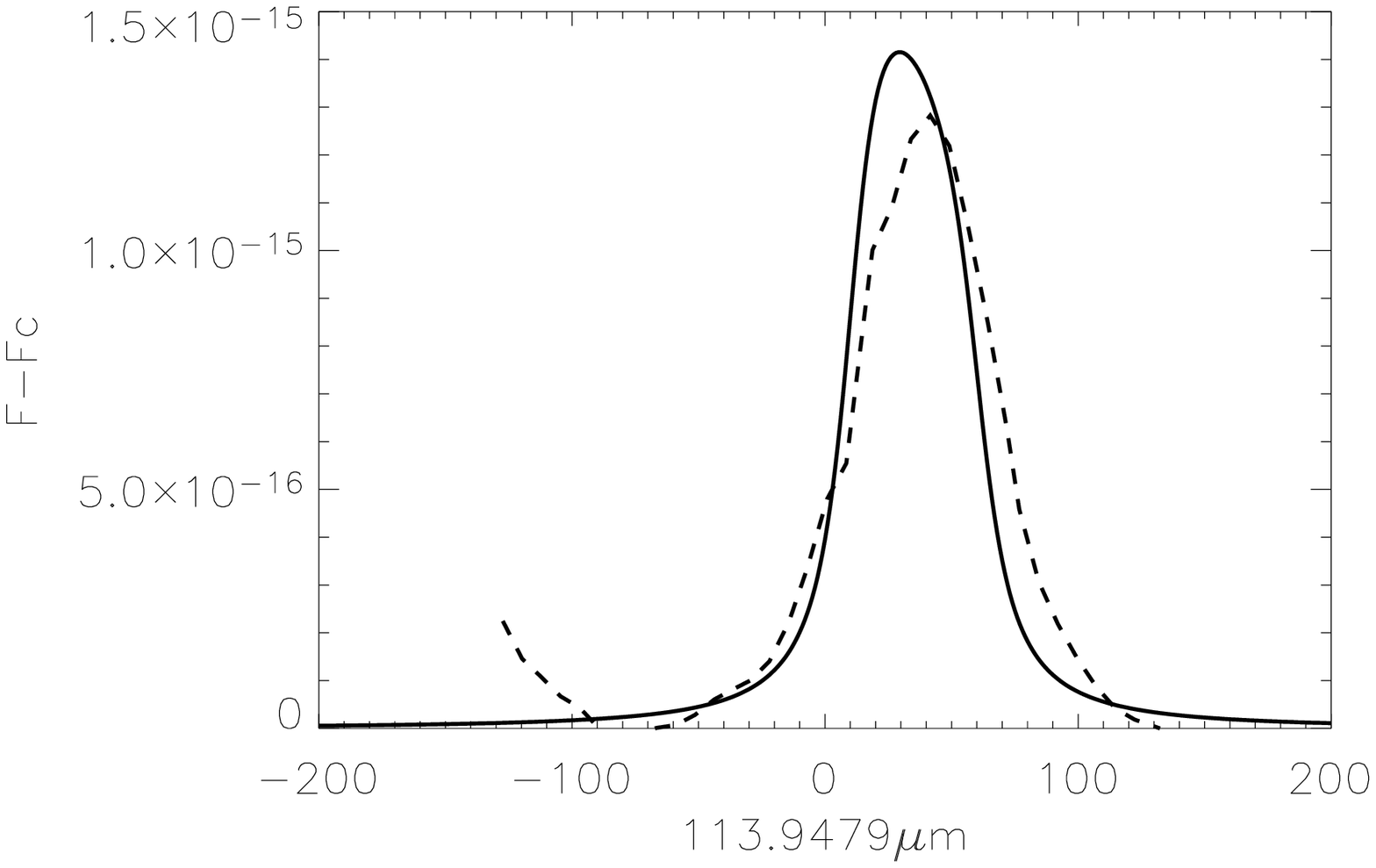}

\caption{Predicted line profiles from the PL2 radiative transfer
model (solid lines), over-plotted on the observed ortho and
para-H$_{2}$O lines (dotted lines). The PL2 model corresponds to
gas of density 3$\times$10$^{5}$cm$^{-3}$, heated up to 300 K and
relaxing to 90-100~K, expanding at a velocity of 30 km~s$^{-1}$
(Table \ref{plat_para}). The ordinate corresponds
to the continuum-subtracted flux and the abscissa is the radial velocity
in km~s$^{-1}$.} \label{agua_results1}
\end{figure}

\begin{landscape}
\begin{table}

\begin{tabular}{c c c c c c c c}
\hline
Wavelength& Transition  & Absorbed flux & Emitted flux & Pred. absorption & Pred. emission & Line peak [1] & Line fit $\chi^{2}$ \\
 ($\mu$m) & &(10$^{-17}$W cm$^{-2}$) &(10$^{-17}$W cm$^{-2}$)& (10$^{-17}$W cm$^{-2}$)& (10$^{-17}$W cm$^{-2}$)& (km s$^{-1}$)& \\
 \hline\hline
 49.281 & p-H$_{2}$O $4_{40}$ -- $3_{31}$ & 0.71 $\pm$ 0.13  & & 0.92& & -9.2 $\pm$ 1.6 & 0.142\\
 56.324 & p-H$_{2}$O $4_{31}$ -- $3_{22}$ & 1.09 $\pm$ 0.11 & 0.59 $\pm$ 0.25 & 1.65 &  & -5.1 $\pm$ 0.5, 44.9 $\pm$ 18.9 & 0.125 \\
 58.698 & o-H$_{2}$O $4_{32}$ -- $3_{21}$ & 1.57 $\pm$ 0.13 & 1.80 $\pm$ 0.14& 1.82&1.95  & -29.9 $\pm$ 2.5, 37.3 $\pm$ 2.9 & 0.0368\\
 61.808 & p-H$_{2}$O $4_{31}$ -- $4_{04}$ & & 0.52 $\pm$ 0.02 &  & 0.85  & 43.5 $\pm$ 1.7 & 0.0728\\
 66.437 & o-H$_{2}$O $3_{30}$ -- $2_{21}$ & 1.51 $\pm$ 0.18 & 1.39 $\pm$ 0.28 & 1.78  & 1.85  & -22.2 $\pm$ 2.6, 40.5 $\pm$ 8.1 & 0.0168\\
 71.066 & p-H$_{2}$O $5_{24}$ -- $4_{13}$ & 0.15 $\pm$ 0.039 & 0.89 $\pm$
 0.059 & &  1.02 & -22.5 $\pm$ 5.8, 28.5 $\pm$ 1.8 & 0.148 \\
 71.539 & p-H$_{2}$O  $7_{17}$ -- $6_{06}$ &  & 1.39 $\pm$ 0.13 &  & 1.02 &
22.3 $\pm$ 2.1 & 0.222\\
 75.380 & o-H$_{2}$O $3_{21}$ -- $2_{12}$  & 0.88 $\pm$ 0.25 & 5.67 $\pm$
0.21& 1.33 & 4.15  & -31.5 $\pm$ 8.9 , 28.5 $\pm$ 1.1 & 0.862\\
78.928 & p-H$_{2}$O $6_{15}$ -- $5_{24}$ & & 0.43 $\pm$ 0.11 & & 0.35    & 16.5 $\pm$ 4.2 & 0.0551\\
82.030 & o-H$_{2}$O $6_{16}$ -- $5_{05}$ & & 2.78 $\pm$ 0.13 & & 1.89    & 22.4 $\pm$ 1.1 & 0.361\\
83.283 & p-H$_{2}$O $6_{06}$ -- $5_{15}$ & & 1.29 $\pm$ 0.08 & & 1.05    & 31.1 $\pm$ 1.8 & 0.181 \\
89.988 & o-H$_{2}$O $3_{22}$ -- $2_{11}$ & & 2.17 $\pm$ 0.17 & & 1.05&     29.1 $\pm$ 2.3 & 0.542\\
94.703 & o-H$_{2}$O $4_{41}$ -- $4_{32}$ & & 0.67 $\pm$ 0.06 & & 1.33    & 29.2 $\pm$ 2.7 & 0.181\\
95.626 & p-H$_{2}$O $5_{15}$ -- $4_{04}$ & &1.42  $\pm$ 0.10 & & 1.25    & 27.4 $\pm$ 1.3 & 0.184\\
95.883 & p-H$_{2}$O $4_{41}$ -- $4_{32}$ & &0.67  $\pm$ 0.06 & &  0.38   & 29.2 $\pm$ 2.7 & 0.121\\
99.492 & o-H$_{2}$O $5_{05}$ -- $4_{14}$ & & 5.61 $\pm$ 0.12 & & 5.25    & 22.2 $\pm$ 0.5 & 0.448\\
100.913 & o-H$_{2}$O $5_{14}$ -- $4_{23}$ & & 3.05 $\pm$ 0.17 & &  2.60   & 21.1 $\pm$ 1.2 & 0.366\\
108.073 & o-H$_{2}$O $2_{21}$ -- $1_{10}$ & & 3.22 $\pm$ 0.08 & &  2.55   & 29.1 $\pm$ 0.7 & 0.418 \\
111.626 & p-H$_{2}$O $5_{24}$ -- $5_{15}$ & & 0.43 $\pm$ 0.03 & &  0.26   & 26.7 $\pm$ 1.7 & 0.068 \\
113.944 & p-H$_{2}$O $5_{33}$ -- $5_{24}$ & & 1.38 $\pm$ 0.06 & &  1.67   & 27.6 $\pm$ 1.3 & 0.207\\
121.719 & o-H$_{2}$O $4_{32}$ -- $4_{23}$ & & 2.28 $\pm$ 0.13 & &  1.55   & 28.1 $\pm$ 1.6 & 0.342 \\
125.354 & p-H$_{2}$O $4_{04}$ -- $3_{13}$ & & 2.25 $\pm$ 0.09 & &  1.98   & 25.5 $\pm$ 1.1 & 0.202\\
126.713 & p-H$_{2}$O $3_{31}$ -- $3_{22}$ & & 0.62 $\pm$ 0.08 & &  0.34   & 30.5 $\pm$ 4.1 & 0.093\\
136.494 & o-H$_{2}$O $3_{30}$ -- $3_{21}$ & & 0.71 $\pm$ 0.05 & &  0.65   & 30.9 $\pm$ 2.3 & 0.113\\
138.527 & p-H$_{2}$O $3_{13}$ -- $2_{02}$ & & 2.67 $\pm$ 0.03 & &   1.66  & 26.4 $\pm$ 0.3 & 0.293\\
144.517 & p-H$_{2}$O $4_{13}$ -- $3_{22}$ & & 1.15 $\pm$ 0.07 &  &  2.38  & 18.4 $\pm$ 0.2 & 0.287\\
146.919 & p-H$_{2}$O $4_{31}$ -- $4_{22}$ & & 1.14 $\pm$ 0.06 &  & 0.98   & 22.1 $\pm$ 1.1 & 0.216\\
156.193 & p-H$_{2}$O $3_{22}$ -- $3_{13}$ & & 1.23 $\pm$ 0.08 & &  1.59   & 32.5 $\pm$ 2.2 & 0.258\\
174.626 & o-H$_{2}$O $3_{03}$ -- $2_{12}$ & & 2.38 $\pm$ 0.19 & &   1.55  & 20.1 $\pm$ 1.6 & 0.333 \\
179.527 & o-H$_{2}$O $2_{12}$ -- $1_{01}$ & & 2.55 $\pm$ 0.31 & &   2.34  & 23.8 $\pm$ 2.9 & 0.204 \\

\hline
\end{tabular}
\caption{H$_{2}$O line fluxes observed and predicted by the
modelling. [1] Observed LSR velocity. Where two values are listed,
these correspond to separate emission peaks.}
\label{aguita_res}
\end{table}

\end{landscape}
Line profile fits from our radiative transfer modelling are shown in
Figure
\ref{agua_results1} and Figure \ref{agua_r2}, while Table
\ref{aguita_res} compares the observed and predicted line fluxes. 
The final column of Table \ref{aguita_res} lists the $\chi^2$ values, 
defined as 
$\chi^2$ = \(\sum_{i=1}^{N}(flux\_mod(i) - flux\_obs(i))^2/N(bins) \),
for the comparison between the modeled (flux\_mod) and observed
(flux\_obs) line profiles.
The line profile fits are excellent for the pure emission lines, while
the lines with P Cygni profiles are reasonably well reproduced.

Overall, our results can
be summarised as follows:
\begin{itemize}
\item Lines arising from energy levels below $\approx$ 560~K are
best reproduced by 70-90~K gas of density
3$\times$10$^{5}$~cm$^{-3}$, expanding at approximately
25-30 km~s$^{-1}$ (\ref{agua_results1}). A graphical representation of the
time evolution of H$_{2}$O, OH and CO is shown in Figure \ref{extended}.
corresponding to Model E2 in Table \ref{exten_para}. Model W2 from
Table \ref{grid_ext} gives very similar results.

\item Lines from higher energy levels, both ortho and para, are
best reproduced by warmer gas of density
3$\times$10$^{5}$~cm$^{-3}$, expanding at approximately 25-30 km
s$^{-1}$, which is heated up to 300 K and then relaxes to 90-100 K
(Figure \ref{agua_r2}). This corresponds to Model PL2 in Table
\ref{plat_para}, which also reproduced the observed CO transitions
with J$_{up}$$\leq$18 (Lerate et al. 2008). Note that this is
simply the best $\chi^2$ fit: models where the shock temperature
is higher, such as PL3, also provide a good fit to the line
profiles. However models where the shock temperature is of the
order of 1000~K, the derived water abundance did not reproduce the
line profile. We should also emphasize that the parameter that
most affected the fits to the water lines was the freeze-out rate,
as this directly affects the fractional abundance of water (since
the higher the freeze-out rate the more oxygen is hydrogenated as
water on the surfaces of the grains before evaporating). 
Based on the investigated models, shock temperatures in the
300-500~K range were found to produce acceptable fits to the
line fluxes and profiles.

\item The PL2 model which fits the higher-excitation lines shown
in Figure \ref{agua_r2} makes a negligible contribution to the
lower-excitation lines shown in Figure \ref{agua_results1},
while the E2 model that fits the lower excitation lines
makes a negligible contribution to the higher-excitation
lines shown in Figure \ref{agua_r2}.

\item The H$_{2}$O to H$_2$ abundance ratio is of the order
of 2-3 $\times$ 10$^{-5}$ for model W2 and 1-5 $\times$ 10$^{-4}$
for model PL2.

 \item For the 14 para-H$_{2}$O lines that show pure emission, 
the ratio of the predicted to observed emission flux is found to be
$0.89\pm0.39$, while for the 10 ortho-H$_{2}$O lines that are
purely in emission this ratio is also $0.89\pm0.39$. We 
interpret this as observational evidence in support of the 
3:1 ortho:para ratio that was adopted for the modeling.

\end{itemize}

\begin{figure}
\centering
\includegraphics[width=4cm,height=4cm]{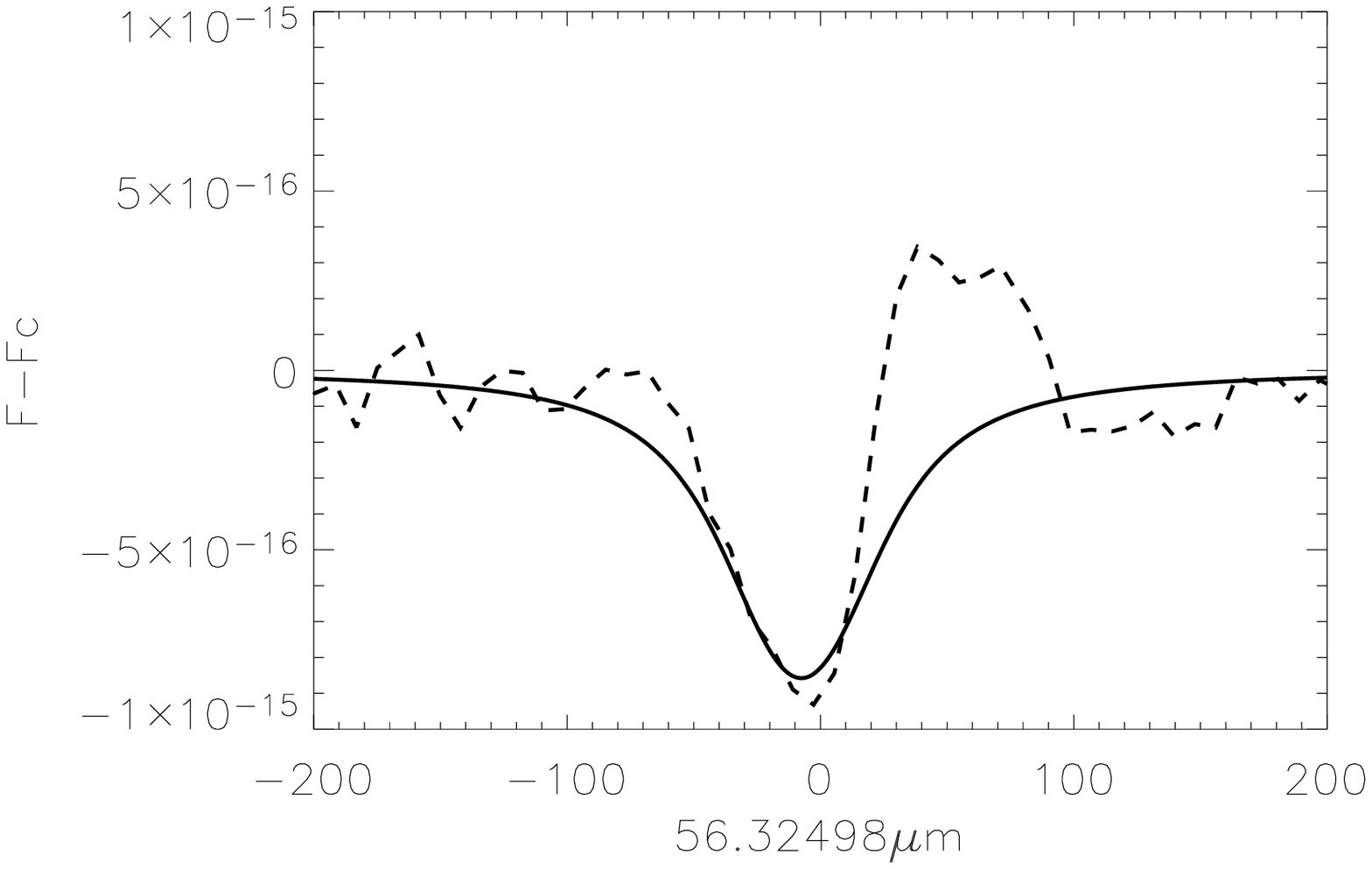}
\includegraphics[width=4cm,height=4cm]{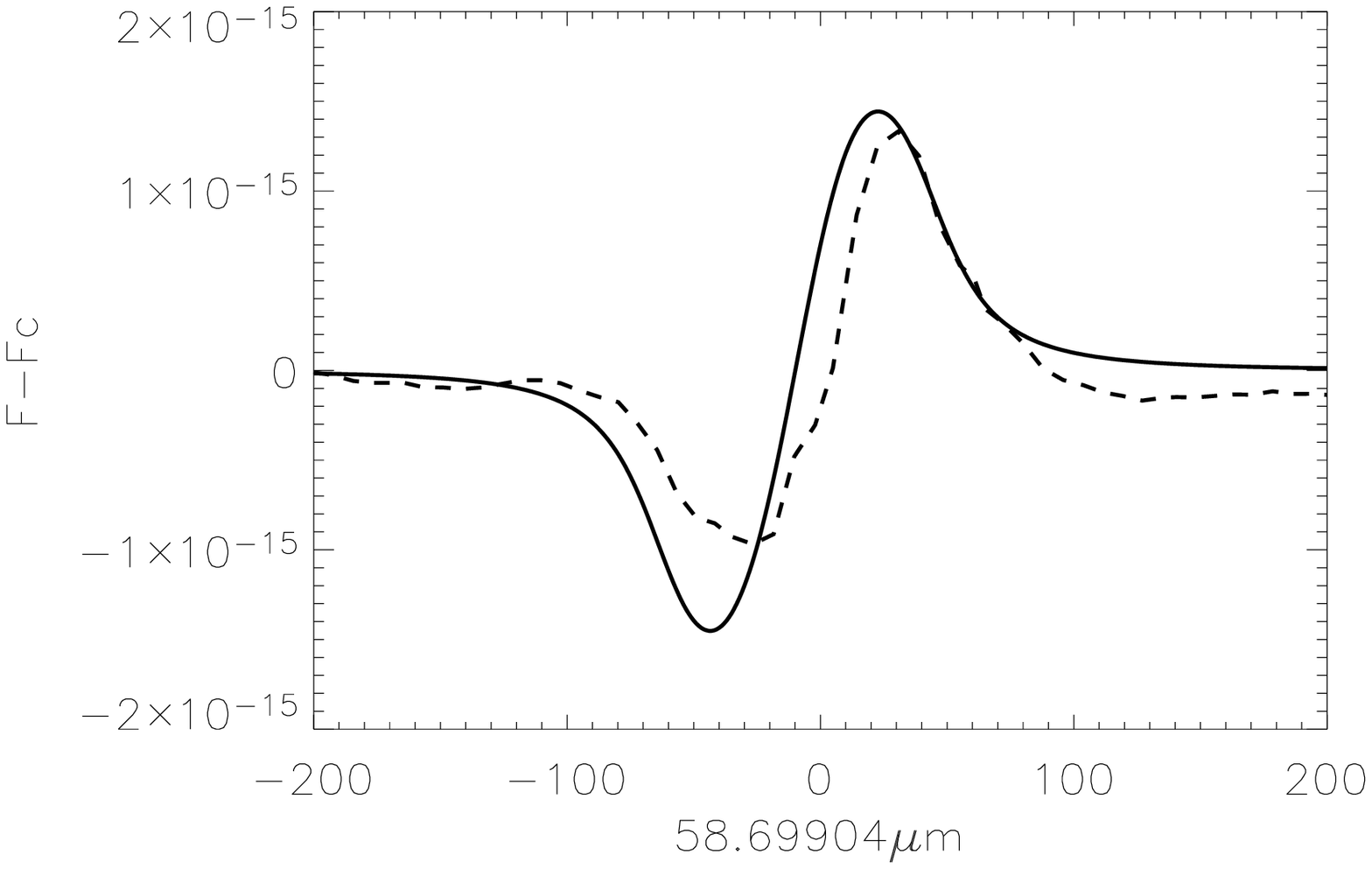}
\includegraphics[width=4cm,height=4cm]{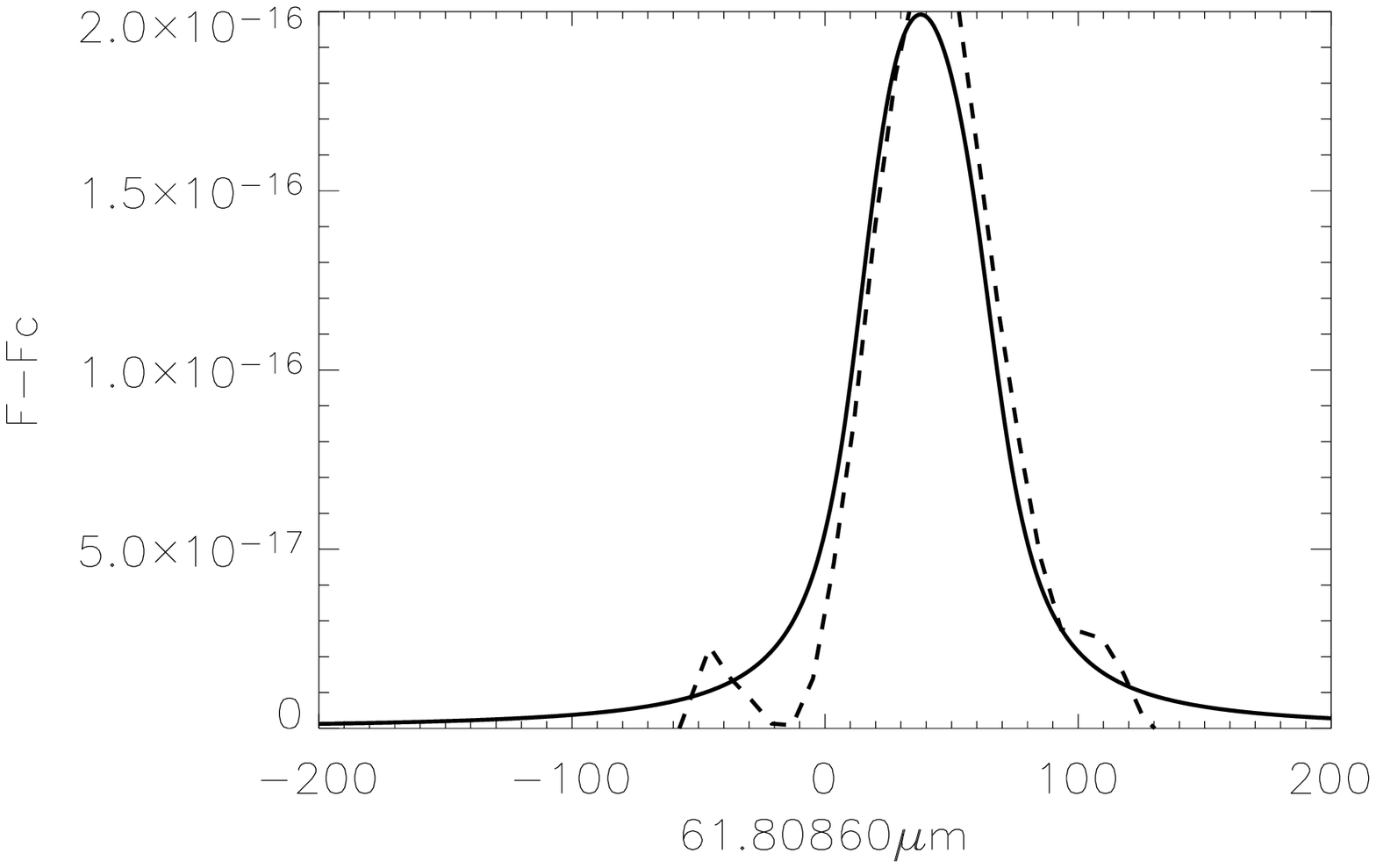}
\includegraphics[width=4cm,height=4cm]{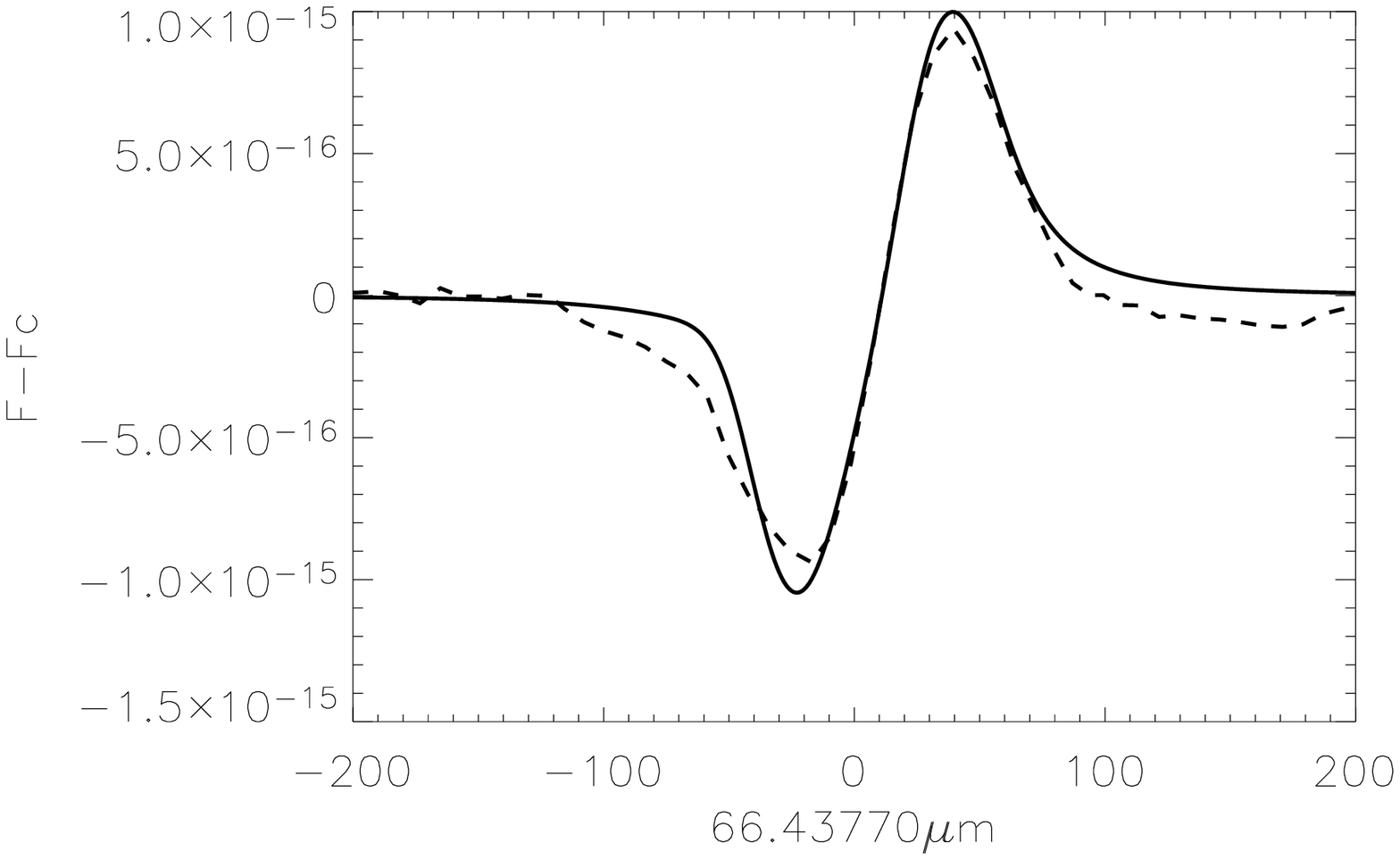}
\includegraphics[width=4cm,height=4cm]{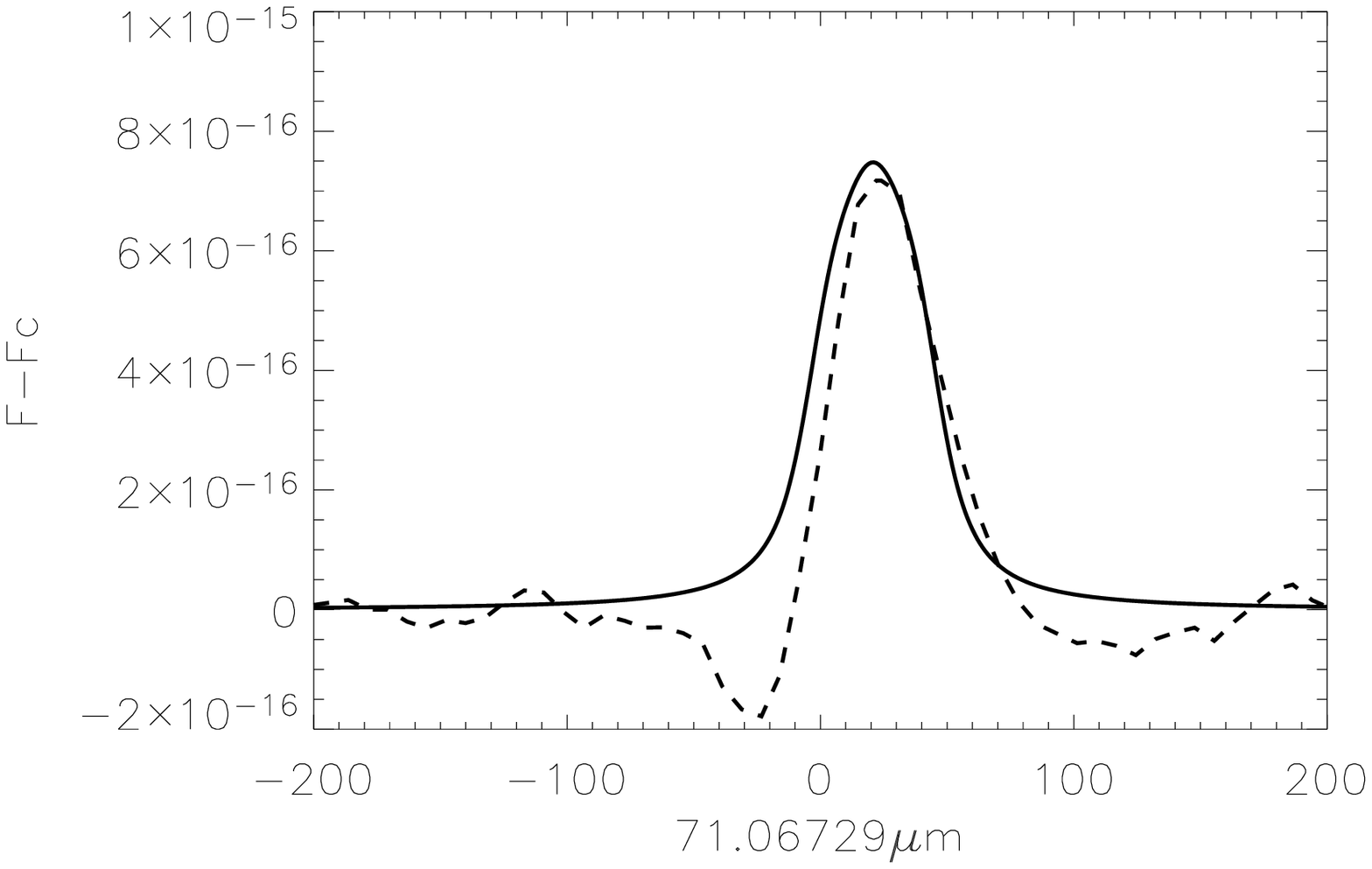}
\includegraphics[width=4cm,height=4cm]{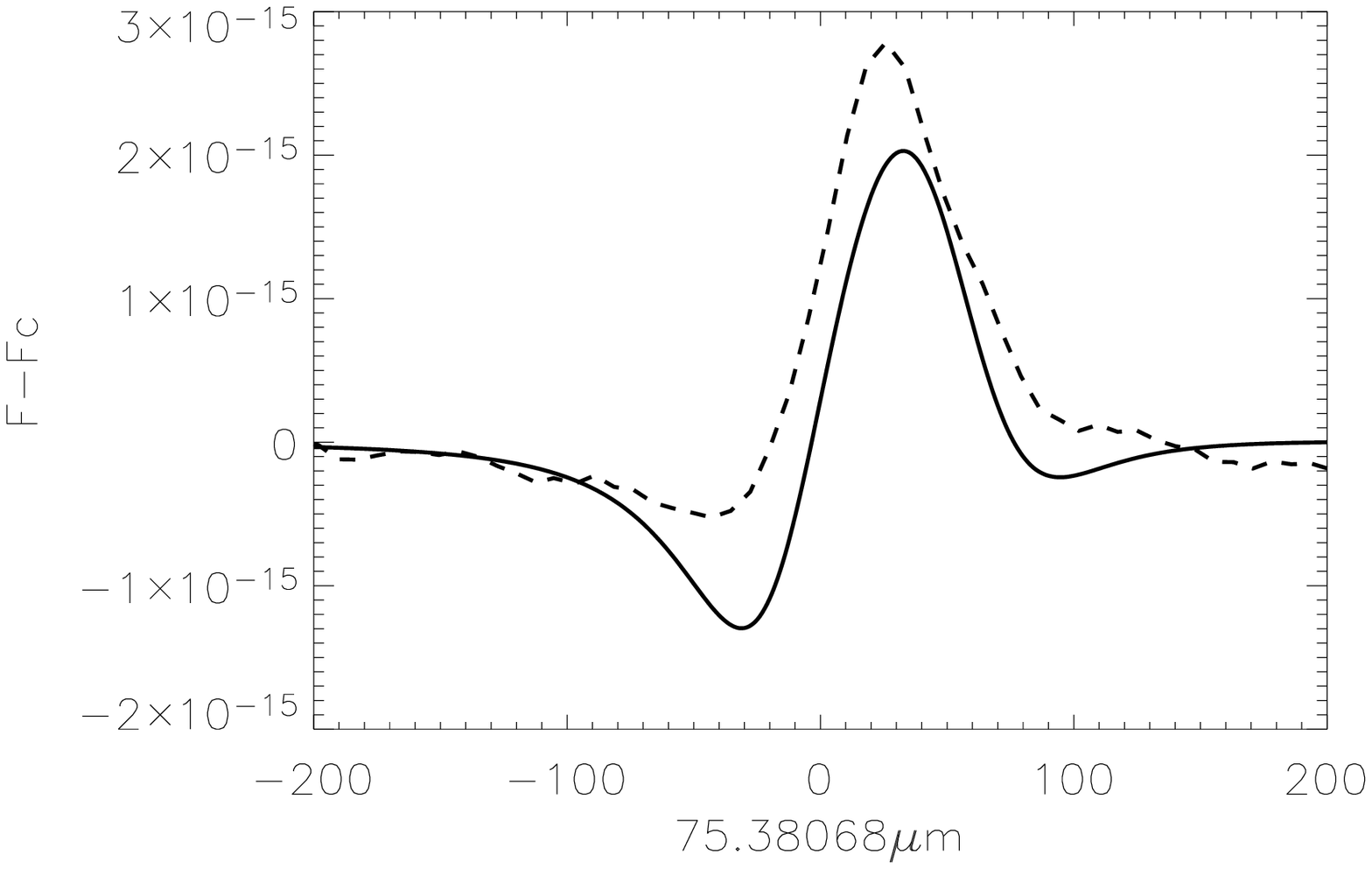}
\includegraphics[width=4cm,height=4cm]{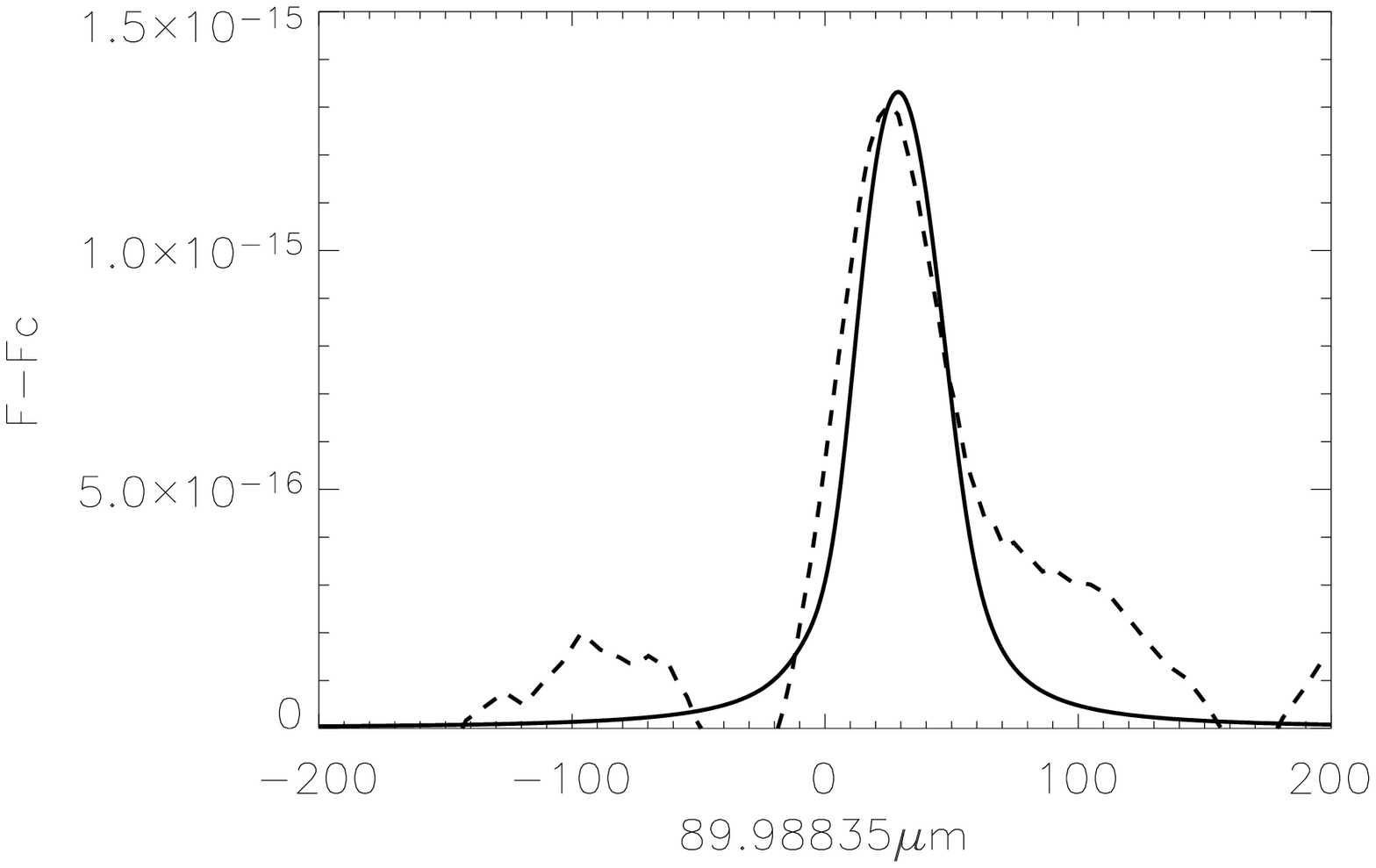}
\includegraphics[width=4cm,height=4cm]{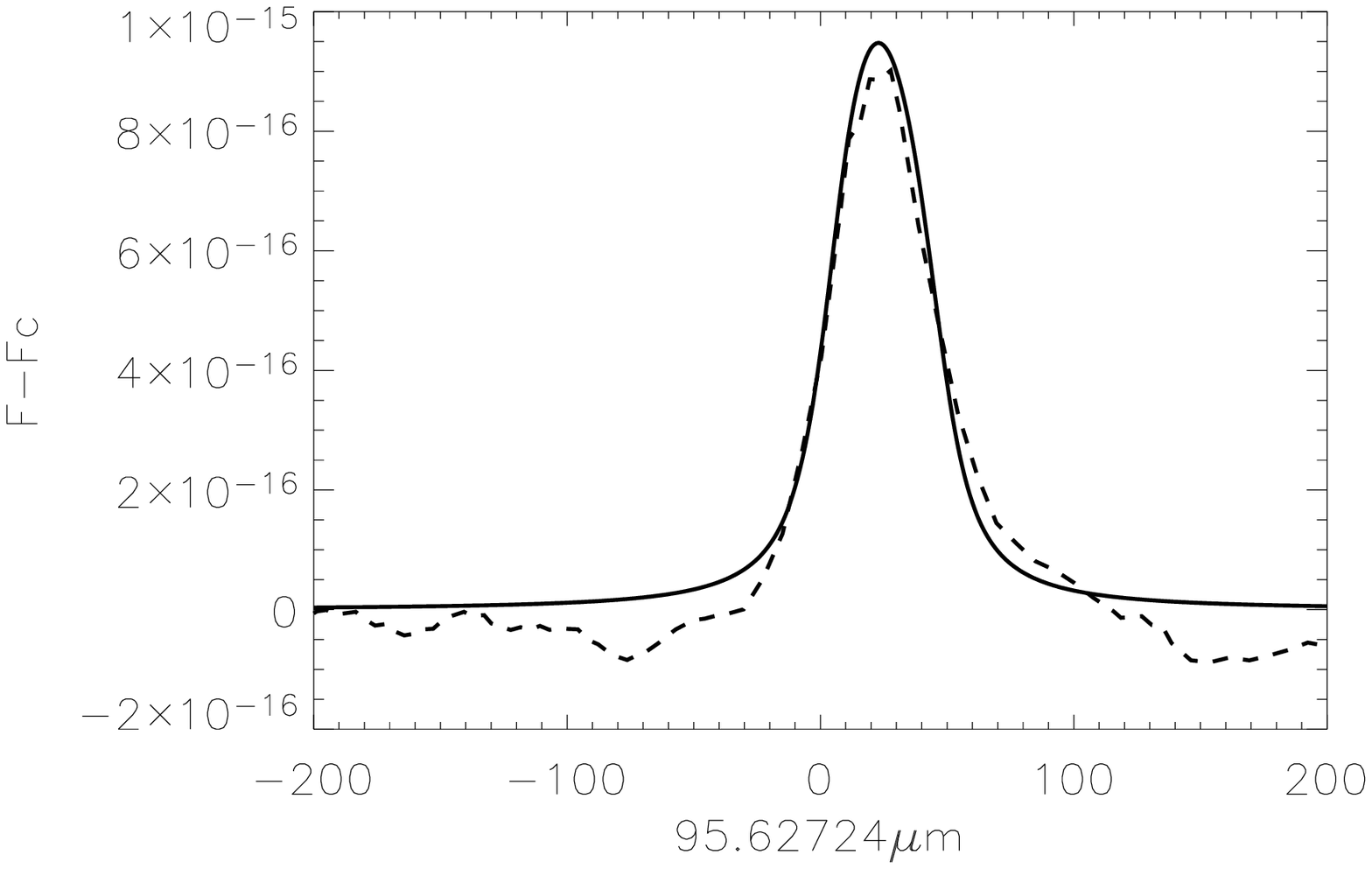}
\includegraphics[width=4cm,height=4cm]{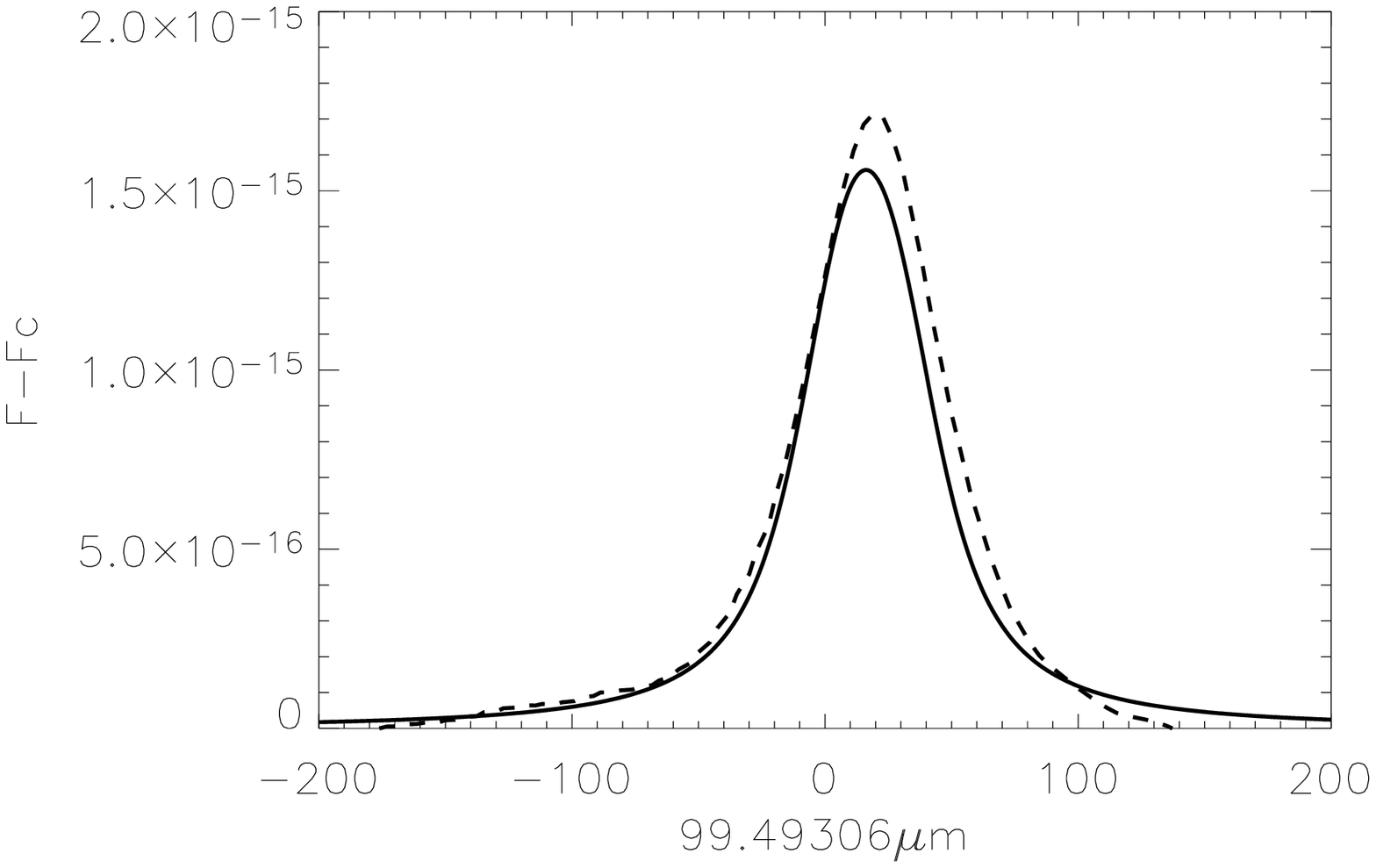}

\caption{Predicted line profiles from the E2 radiative transfer
model (solid lines) over-plotted on the observed ortho and
para-H$_{2}$O lines (dotted lines). The E2 model corresponds to
gas of density 3$\times$10$^{5}$cm$^{-3}$ at 70-90~K
expanding at a velocity of 30 km s$^{-1}$ (Table
\ref{exten_para}). The ordinate corresponds to the
continuum-subtracted flux and the abscissa is the radial velocity in
km~s$^{-1}$.} \label{agua_r2}

\end{figure}

\addtocounter{figure}{-1}
\begin{figure}
    \centering
\includegraphics[width=4cm,height=4cm]{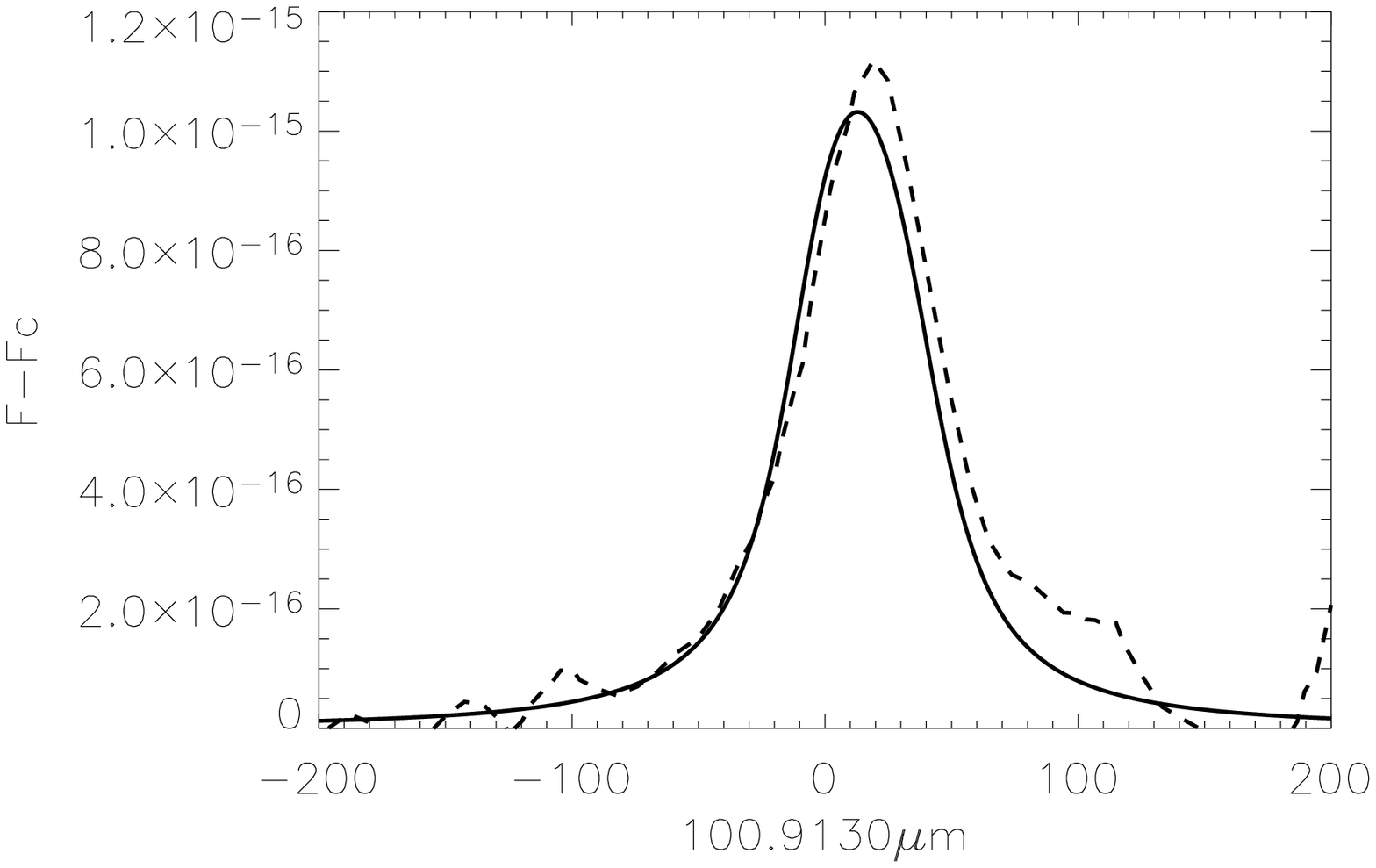}
\includegraphics[width=4cm,height=4cm]{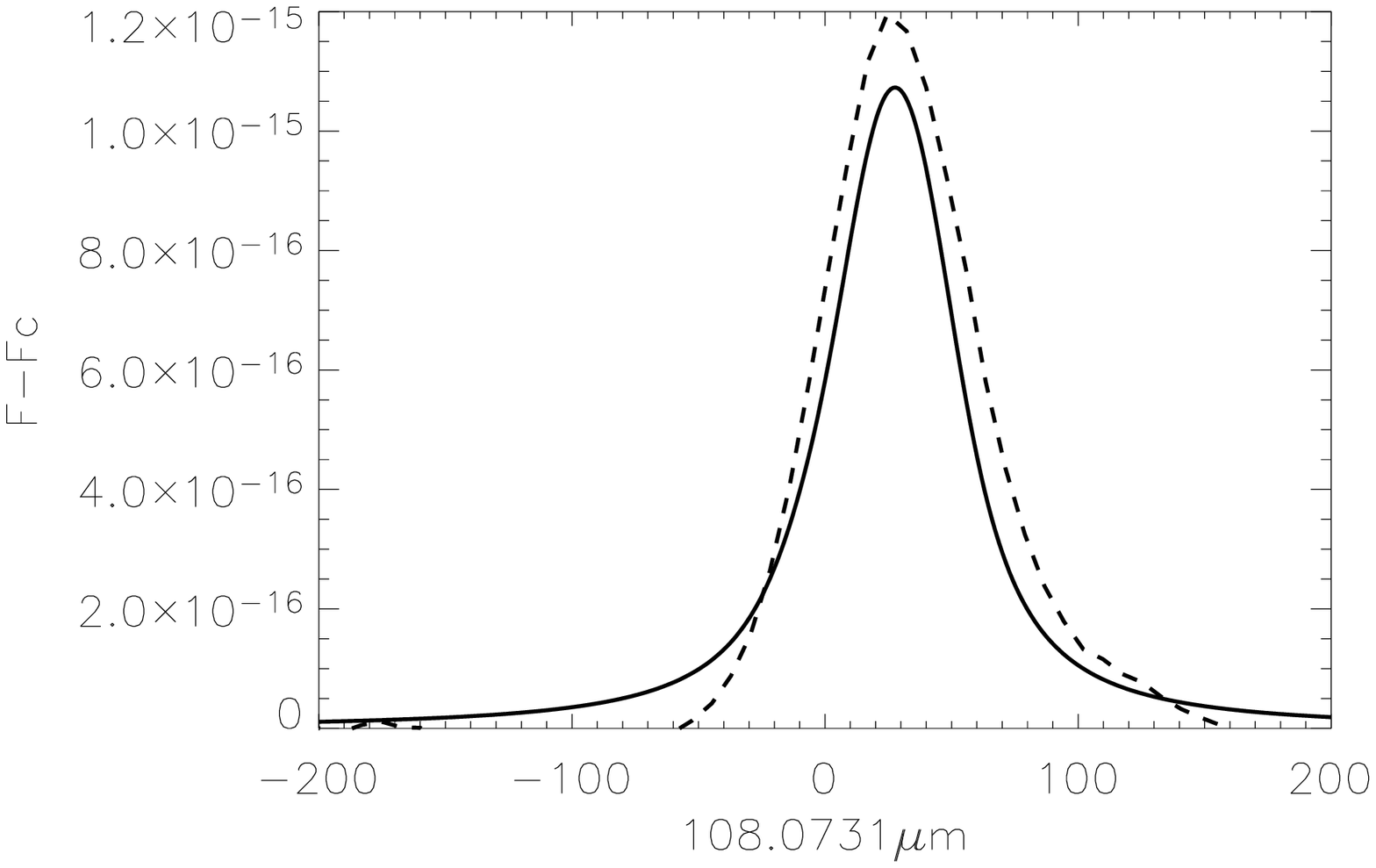}
\includegraphics[width=4cm,height=4cm]{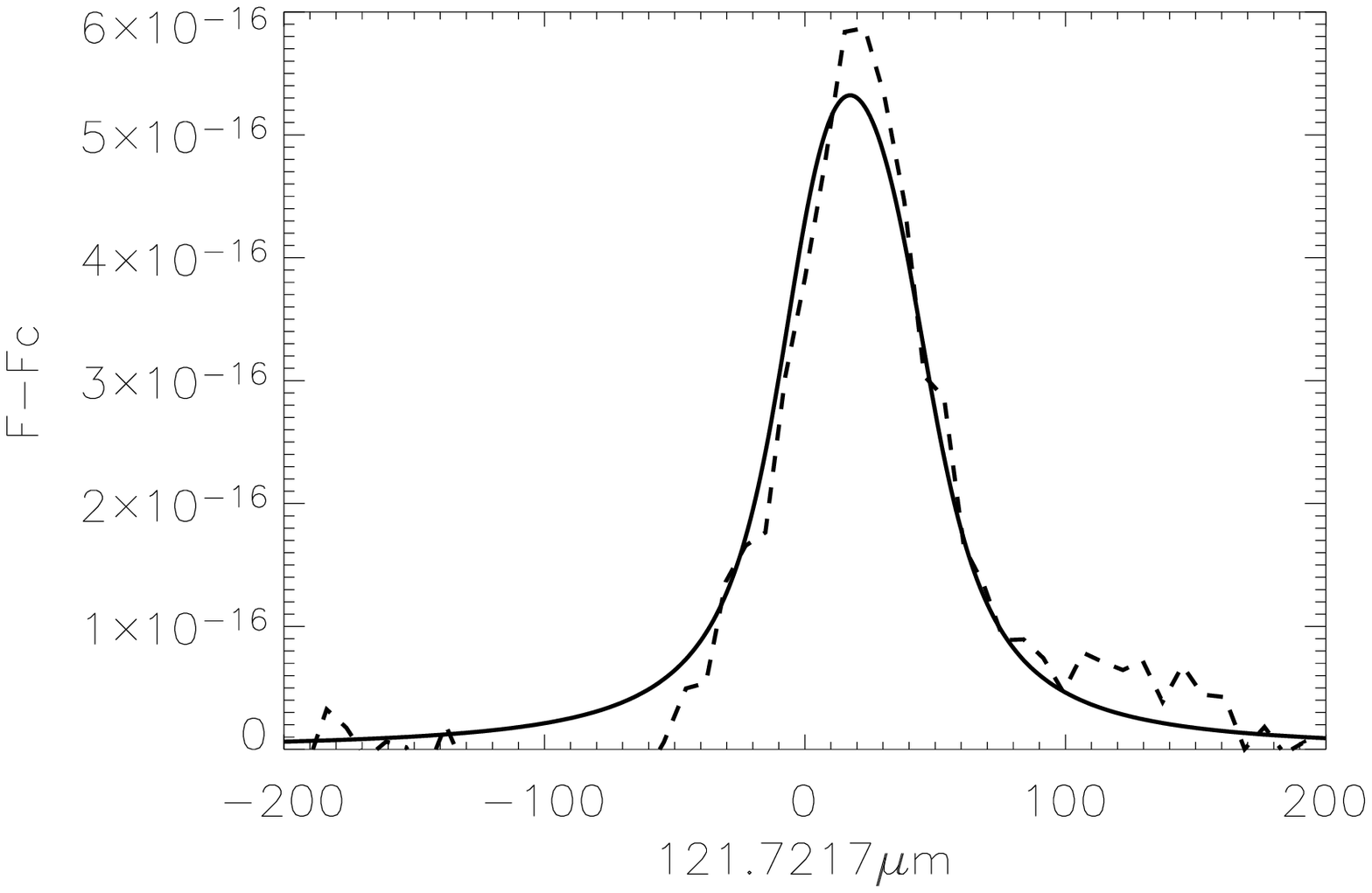}
\includegraphics[width=4cm,height=4cm]{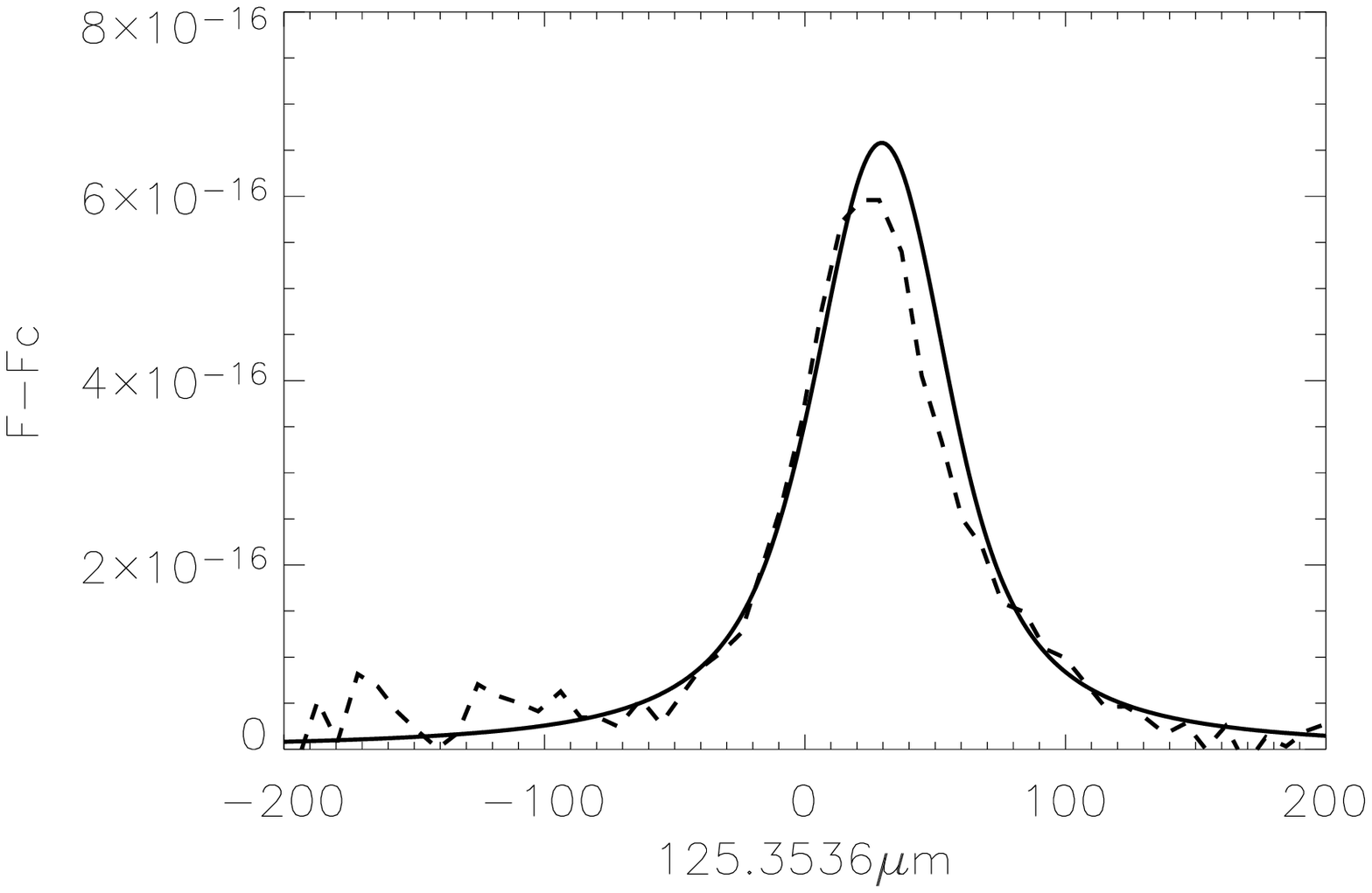}
\includegraphics[width=4cm,height=4cm]{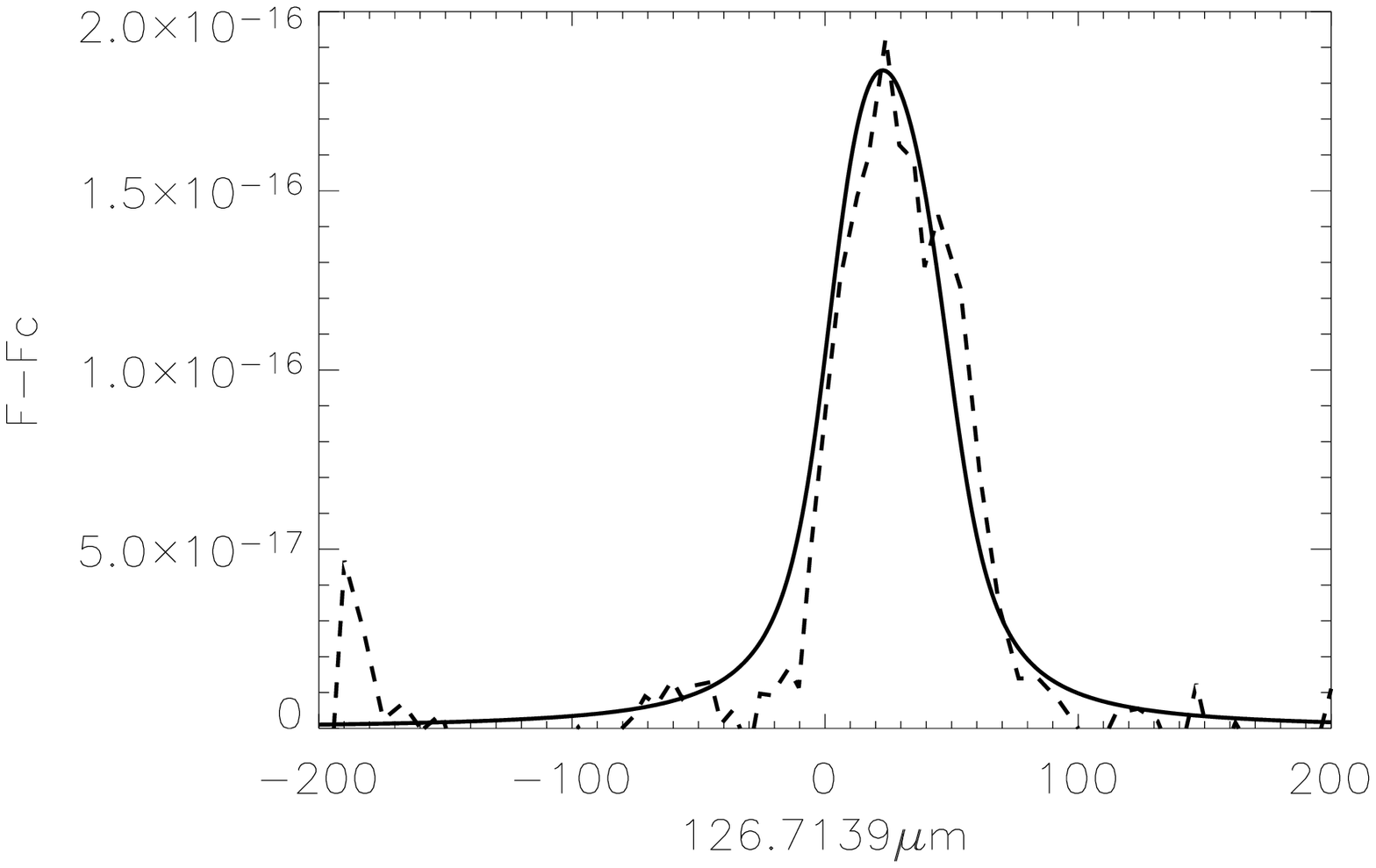}
\includegraphics[width=4cm,height=4cm]{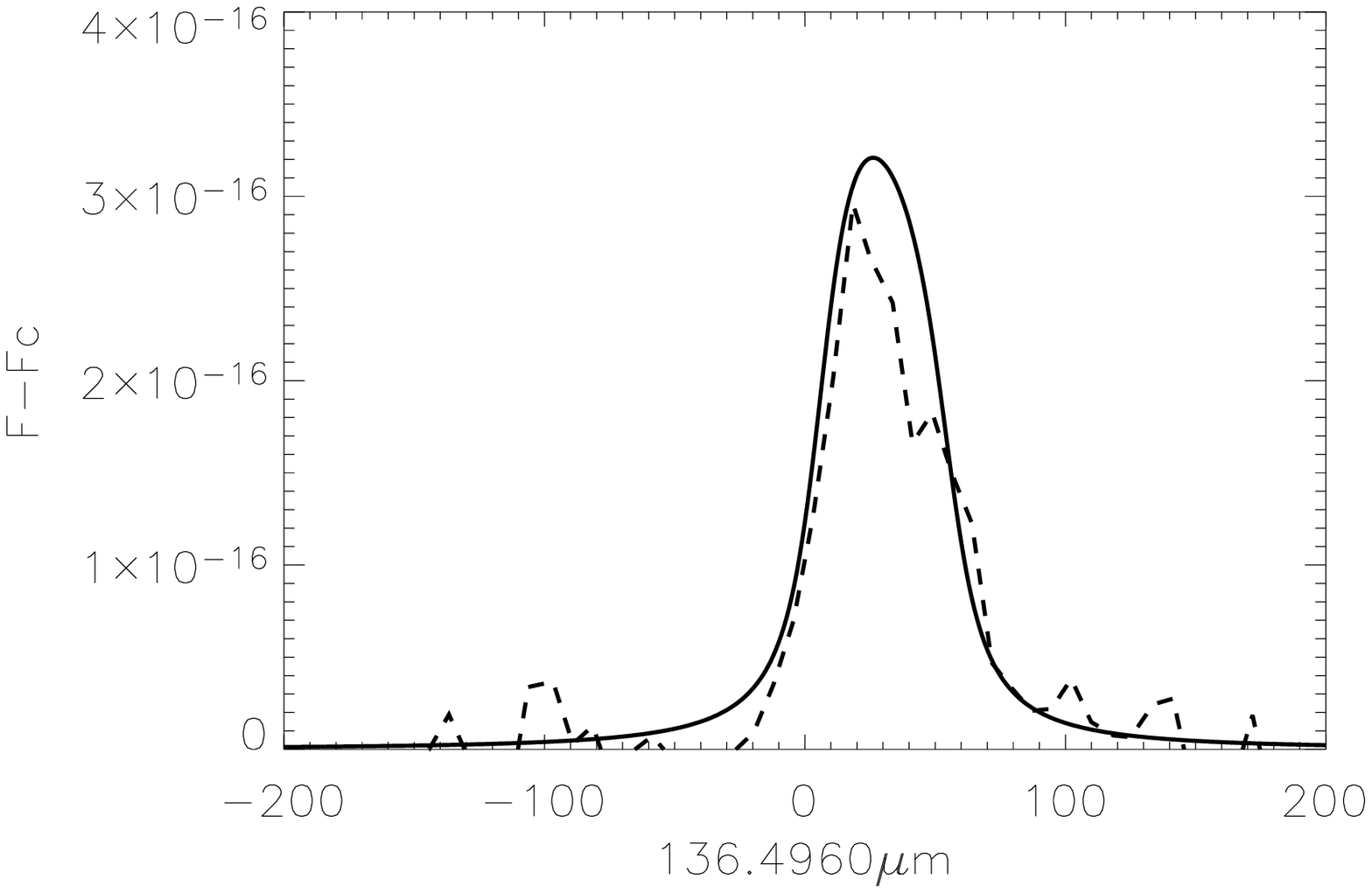}
\includegraphics[width=4cm,height=4cm]{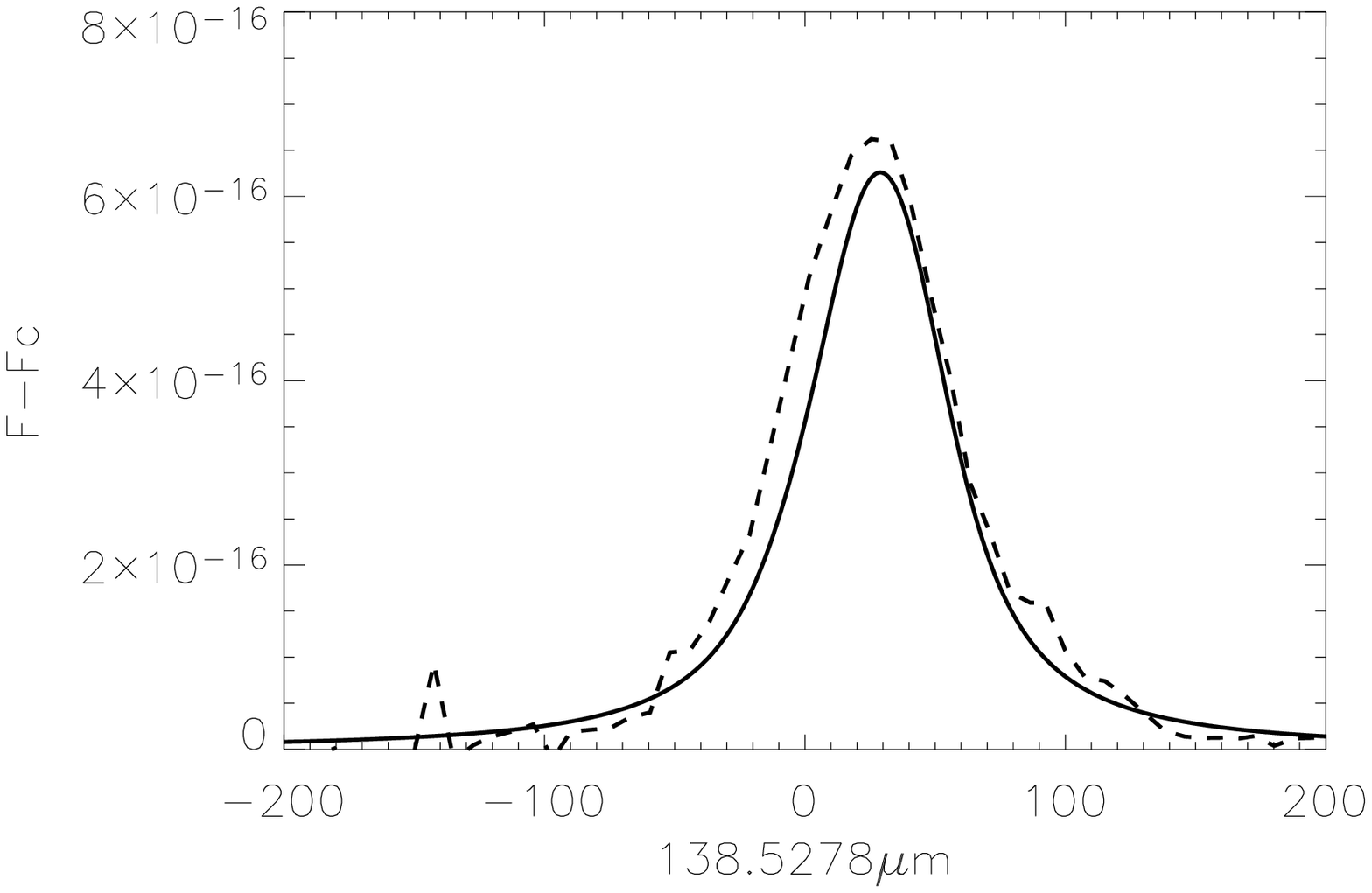}
\includegraphics[width=4cm,height=4cm]{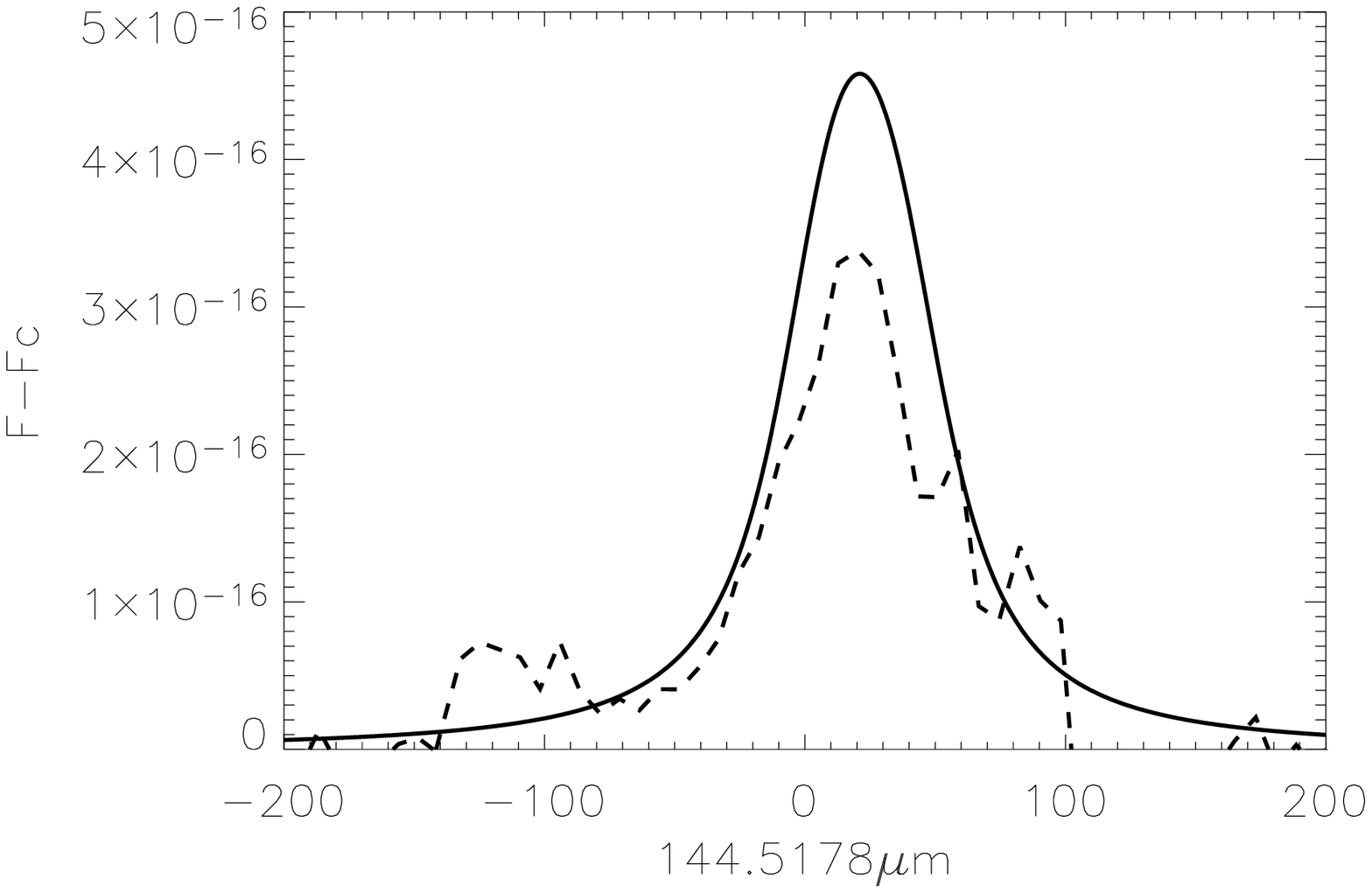}
\includegraphics[width=4cm,height=4cm]{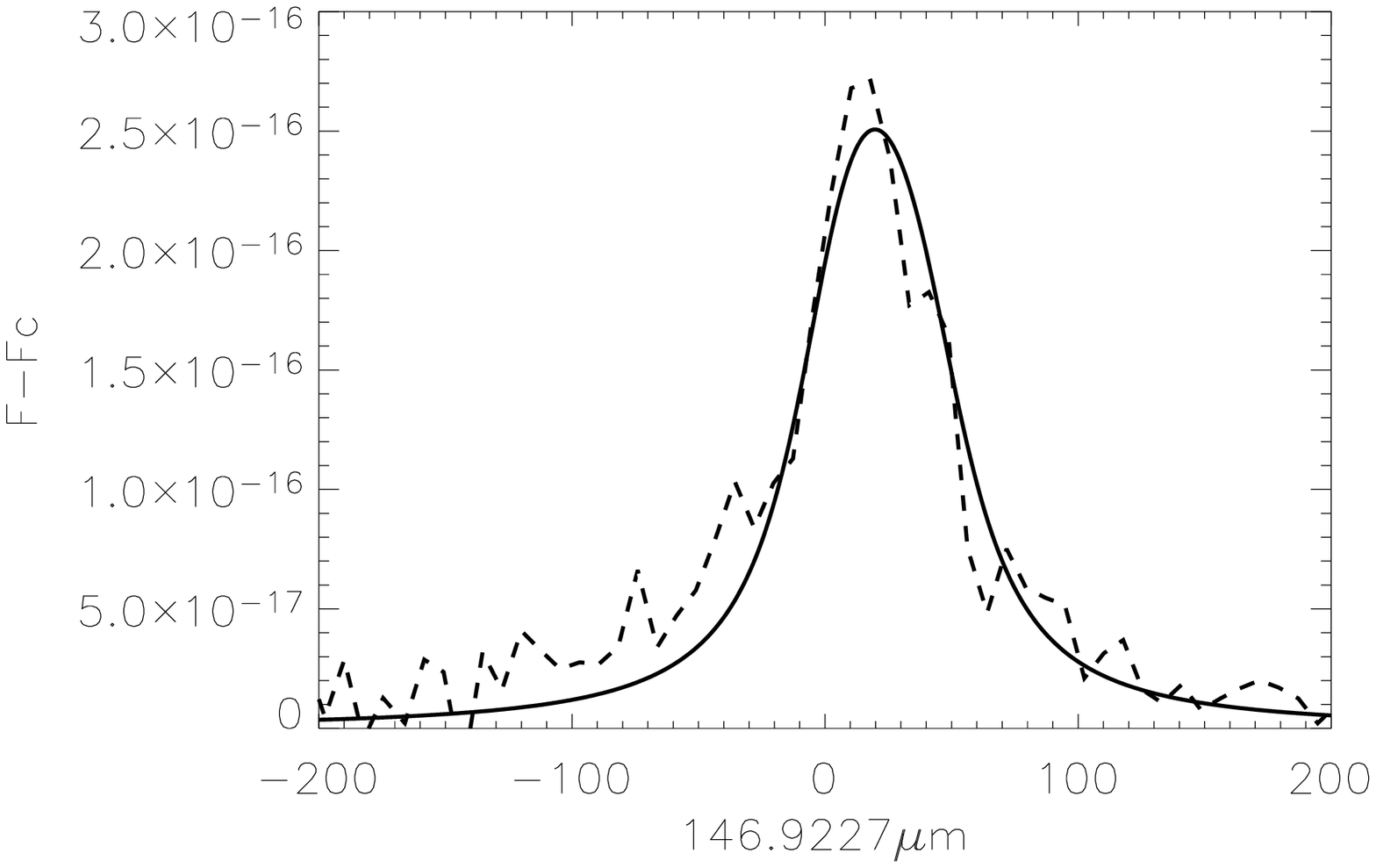}
\includegraphics[width=4cm,height=4cm]{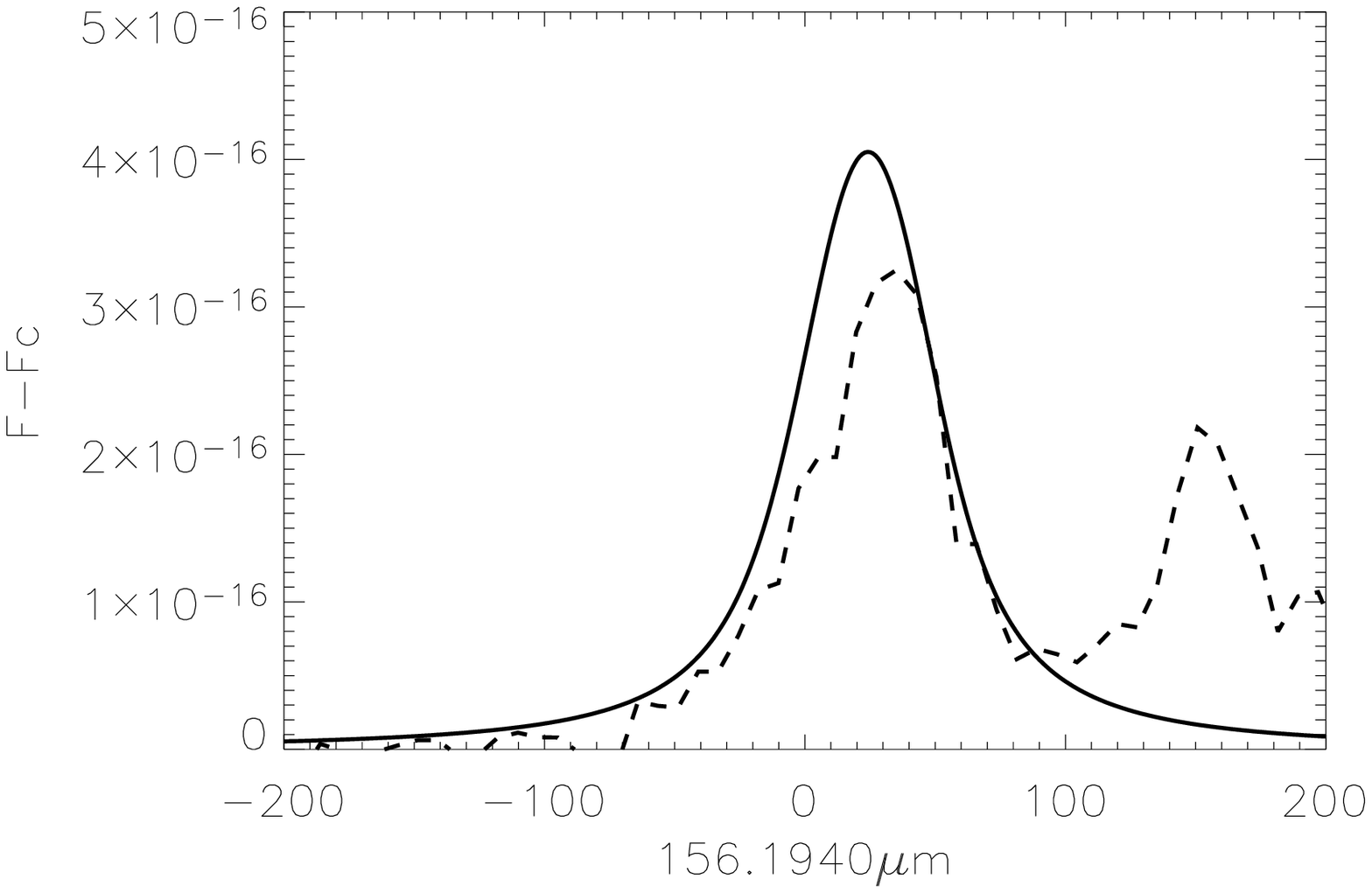}
\includegraphics[width=4cm,height=4cm]{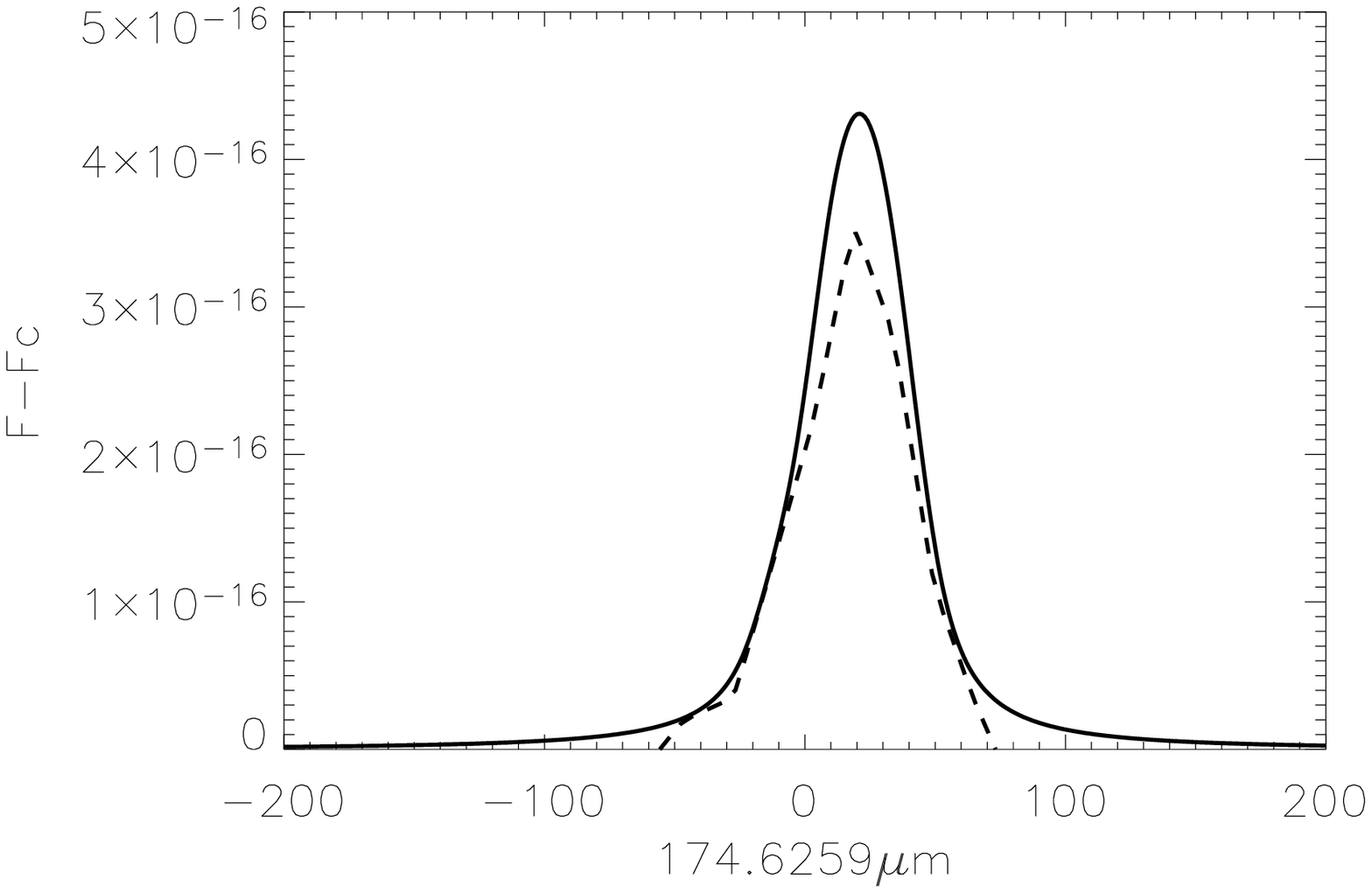}
\includegraphics[width=4cm,height=4cm]{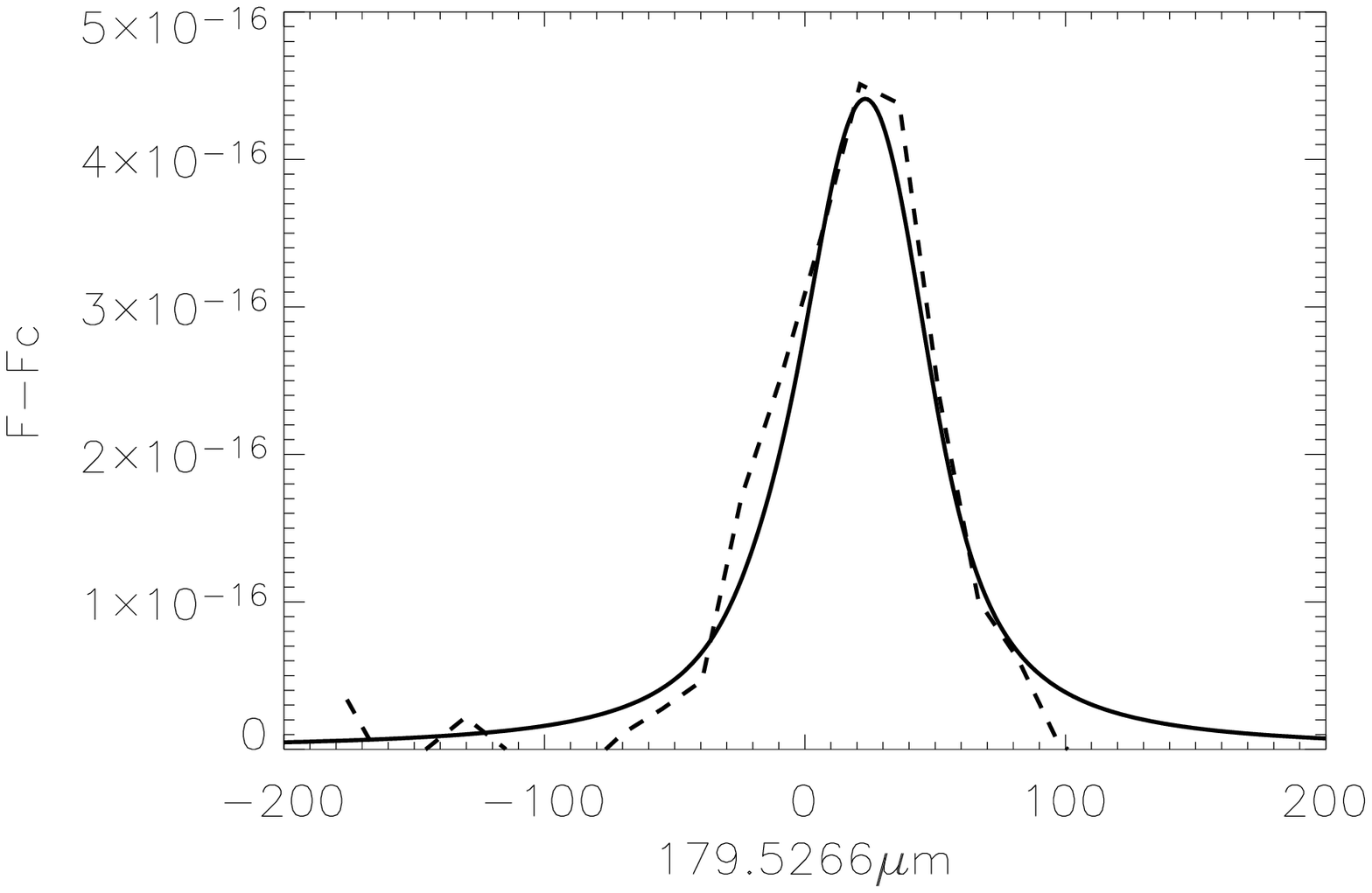}

\caption{Continued}
\end{figure}


The results from our modeling of the observed
far-IR ortho and para-H$_{2}$O lines from Orion-KL show that two
different chemical models are needed to reproduce the H$_{2}$O lines
in the LWS wavelength range. For lines arising from energy levels
below $\approx$ 560~K, our modelling results are in agreement with
the findings of Cernicharo et al. (2006); our chemical model of an
expanding gas with velocity of 30 km~s$^{-1}$ at 70-90~K is able to
reproduce the line fluxes and line profiles of both the ortho- and
para-H$_{2}$O lines, with an H$_{2}$O/H$_{2}$ abundance ratio of the
order of 2-3 $\times$ 10$^{-5}$. For the water lines that exhibit
P~Cygni profiles, our current profile fits appear to provide
an improvement when compared to the fits presented by Cernicharo et al.
(2006). \\

However, for lines arising from higher energy levels (above 560~K)
the model that best reproduces both the H$_{2}$O line fluxes and
their profiles is of warmer gas which is initially heated up to
$\sim$300~K and then relaxes to 90-100 K. The corresponding
H$_{2}$O/H$_{2}$ abundance ratio depends on the time stage within
the PL2 model but is of the order of 1-5 $\times$ 10$^{-4}$,
within reach of the value of 5 $\times$ 10$^{-4}$ derived by
Harwit et al. (1998) from their modelling of eight Orion-KL LWS-FP
water lines. For the seven water lines in common, Harwit et al.'s
mean observed/predicted line flux ratio was 1.9$\pm$1.5, versus a
mean value of 1.4$\pm$0.4 from our modelling. Our PL2 model was
also found to reproduce well the observed LWS-FP CO transitions
having J$_{up}$$\leq$18, interpreted as arising from the Plateau
region within the extended warm component emission (Lerate et al.
2008). Note, however, that, for higher-J CO lines our work shows
that a higher temperature gas is needed, in agreement with other
authors (e.g. Watson et al. 1985), confirming the findings of
Lerate et al. (2008) that the observed molecular emission arises
from multiple components that differ in density and temperature.
  \\

To conclude, we find that, taken together, Plateau region model PL2
and extended-gas region model E2 can match all of the measured far-IR
water lines from Orion-KL. The main difference between the two modelled
zones is that the Plateau region has warmer temperatures, with a
consequent impact on the chemical evolution. As noted by Cernicharo et al.
(2006), radiative pumping due to the strong IR dust continuum radiation
field is sufficient to populate the higher excitation rotational water
lines. The H$_2$O/H$_2$ ratio in the extended-gas region is
found to be $\sim (2-3)\times10^{-5}$, similar to the value found by
Cernicharo et al. (1996) from their line modelling, but we find
a significantly higher ratio, $\sim (1-5)\times10^{-4}$,
in our Plateau region models that fit the profiles and fluxes of the
higher-excitation water lines.

Our present results should be taken together with those of Lerate
et al. (2008), who analysed ISO LWS-FP observations of multiple
rotational lines of CO and found that in order to explain the
emission from all of the the CO transitions, hot cores as well as
shocked regions (with temperatures ranging from 300 to 1000~K) had
to be present within the ISO beam.

\section*{Acknowledgments}
      This work made use of the Miracle Supercomputer, at the HiPerSPACE Computing Centre,
      UCL, which was funded by the U.K. Particle Physics and Astronomy
      Research Council.
      The {\em ISO} Spectral Analysis Package (ISAP) is a joint
      development by the LWS and SWS Instrument Teams and Data
      Centres. Contributing institutes are CESR, IAS, IPAC, MPE,
      RAL and SRON. LIA ia a joint development of the {\em ISO}-LWS
      Instrument Team at Rutherford Appleton Laboratories (RAL,
      UK- the PI institute) and the Infrared Processing and
      Analysis Center (IPAC/Caltech, USA).\\

\end{document}